\documentclass[superscriptaddress,reprint]{revtex4-2}
\usepackage{amsmath,amssymb,amsfonts}
\usepackage{hyperref}
\usepackage{graphicx}
\usepackage{times}
\usepackage[normalem]{ulem}
\usepackage{siunitx}

\usepackage{enumitem}
\begin{document}

	\title[Article Title]{Robust quantum computational advantage with programmable \\ 3050-photon Gaussian boson sampling}

        \author{Hua-Liang Liu}
        \altaffiliation{These authors contributed equally to this work.}
    \affiliation{Hefei National Laboratory for Physical Sciences at Microscale and School of Physical Sciences,New Cornerstone Science Laboratory, University of Science and Technology of China, Hefei, Anhui, 230026, China}
    \affiliation{CAS Centre for Excellence and Synergetic Innovation Centre in Quantum Information and Quantum Physics, University of Science and Technology of China, Shanghai, 201315, China}
    \affiliation{Hefei National Laboratory, University of Science and Technology of China, Hefei 230088, China}

            \author{Hao Su}
            \altaffiliation{These authors contributed equally to this work.}
    \affiliation{Hefei National Laboratory for Physical Sciences at Microscale and School of Physical Sciences,New Cornerstone Science Laboratory, University of Science and Technology of China, Hefei, Anhui, 230026, China}
    \affiliation{CAS Centre for Excellence and Synergetic Innovation Centre in Quantum Information and Quantum Physics, University of Science and Technology of China, Shanghai, 201315, China}
    \affiliation{Hefei National Laboratory, University of Science and Technology of China, Hefei 230088, China}

                \author{Si-Qiu Gong}
                \altaffiliation{These authors contributed equally to this work.}
    \affiliation{Hefei National Laboratory for Physical Sciences at Microscale and School of Physical Sciences,New Cornerstone Science Laboratory, University of Science and Technology of China, Hefei, Anhui, 230026, China}
    \affiliation{CAS Centre for Excellence and Synergetic Innovation Centre in Quantum Information and Quantum Physics, University of Science and Technology of China, Shanghai, 201315, China}
    \affiliation{Hefei National Laboratory, University of Science and Technology of China, Hefei 230088, China}

                    \author{Yi-Chao Gu}
    \affiliation{Hefei National Laboratory for Physical Sciences at Microscale and School of Physical Sciences,New Cornerstone Science Laboratory, University of Science and Technology of China, Hefei, Anhui, 230026, China}
    \affiliation{CAS Centre for Excellence and Synergetic Innovation Centre in Quantum Information and Quantum Physics, University of Science and Technology of China, Shanghai, 201315, China}
    \affiliation{Hefei National Laboratory, University of Science and Technology of China, Hefei 230088, China}

                        \author{Hao-Yang Tang}
    \affiliation{Hefei National Laboratory for Physical Sciences at Microscale and School of Physical Sciences,New Cornerstone Science Laboratory, University of Science and Technology of China, Hefei, Anhui, 230026, China}
    \affiliation{CAS Centre for Excellence and Synergetic Innovation Centre in Quantum Information and Quantum Physics, University of Science and Technology of China, Shanghai, 201315, China}
    \affiliation{Hefei National Laboratory, University of Science and Technology of China, Hefei 230088, China}

                            \author{Meng-Hao Jia}
    \affiliation{Hefei National Laboratory for Physical Sciences at Microscale and School of Physical Sciences,New Cornerstone Science Laboratory, University of Science and Technology of China, Hefei, Anhui, 230026, China}
    \affiliation{CAS Centre for Excellence and Synergetic Innovation Centre in Quantum Information and Quantum Physics, University of Science and Technology of China, Shanghai, 201315, China}
    \affiliation{Hefei National Laboratory, University of Science and Technology of China, Hefei 230088, China}

                                \author{Qian Wei}
    \affiliation{Hefei National Laboratory for Physical Sciences at Microscale and School of Physical Sciences,New Cornerstone Science Laboratory, University of Science and Technology of China, Hefei, Anhui, 230026, China}
    \affiliation{CAS Centre for Excellence and Synergetic Innovation Centre in Quantum Information and Quantum Physics, University of Science and Technology of China, Shanghai, 201315, China}
    \affiliation{Hefei National Laboratory, University of Science and Technology of China, Hefei 230088, China}

        \author{Yukun Song}
                            \affiliation{Jinan Institute of Quantum Technology and Hefei National Laboratory Jinan Branch, Jinan 250101, China}

    \author{Dongzhou Wang}
                            \affiliation{Jinan Institute of Quantum Technology and Hefei National Laboratory Jinan Branch, Jinan 250101, China}

                                                        \author{Mingyang Zheng}
                            \affiliation{Jinan Institute of Quantum Technology and Hefei National Laboratory Jinan Branch, Jinan 250101, China}

                                            \author{Faxi Chen}
                            \affiliation{Jinan Institute of Quantum Technology and Hefei National Laboratory Jinan Branch, Jinan 250101, China}

                                                                                    \author{Libo Li}
                            \affiliation{Jinan Institute of Quantum Technology and Hefei National Laboratory Jinan Branch, Jinan 250101, China}

                                                        \author{Siyu Ren}
                            \affiliation{State Key Laboratory of Quantum Optics Technologies and Devices, Institute of Opto-Electronics, Shanxi University, Taiyuan 030006, China}

                                                        \author{Xuezhi Zhu}
                            \affiliation{State Key Laboratory of Quantum Optics Technologies and Devices, Institute of Opto-Electronics, Shanxi University, Taiyuan 030006, China}

                                                                                    \author{Meihong Wang}
                            \affiliation{State Key Laboratory of Quantum Optics Technologies and Devices, Institute of Opto-Electronics, Shanxi University, Taiyuan 030006, China}

    \author{Yaojian Chen}
    \affiliation{Department of Computer Science and Technology and
 Beijing National Research Center for Information Science and Technology,
 Tsinghua University, Beijing 100084, China}

        \author{Yanfei Liu}
    \affiliation{National supercomputer center in wuxi, Wuxi 214026, China}

     \author{Longsheng Song}
    \affiliation{School of Computer Science and Technology, University of Science and Technology of China, Hefei, Anhui, 230026, China}

         \author{Pengyu Yang}
    \affiliation{School of Computer Science and Technology, University of Science and Technology of China, Hefei, Anhui, 230026, China}

      \author{Junshi Chen}
    \affiliation{School of Computer Science and Technology, University of Science and Technology of China, Hefei, Anhui, 230026, China}
    \affiliation{Laoshan Laboratory, Qingdao 266237, China}

                                      \author{Hong An}
    \affiliation{School of Computer Science and Technology, University of Science and Technology of China, Hefei, Anhui, 230026, China}
    \affiliation{Laoshan Laboratory, Qingdao 266237, China}

    \author{Lei Zhang}
    \affiliation{Shanghai Precilasers Technology Co.Ltd., Shanghai 201822, China}

                                \author{Lin Gan}
    \affiliation{Department of Computer Science and Technology and
 Beijing National Research Center for Information Science and Technology,
 Tsinghua University, Beijing 100084, China}

     \author{Guangwen Yang}
    \affiliation{Department of Computer Science and Technology and
 Beijing National Research Center for Information Science and Technology,
 Tsinghua University, Beijing 100084, China}

        \author{Jia-Min Xu}
    \affiliation{Jiuzhang Quantum Technology Co.Ltd., Jinan 250102, China}

                                                                     \author{Yu-Ming He}
    \affiliation{Hefei National Laboratory for Physical Sciences at Microscale and School of Physical Sciences,New Cornerstone Science Laboratory, University of Science and Technology of China, Hefei, Anhui, 230026, China}
    \affiliation{CAS Centre for Excellence and Synergetic Innovation Centre in Quantum Information and Quantum Physics, University of Science and Technology of China, Shanghai, 201315, China}
    \affiliation{Hefei National Laboratory, University of Science and Technology of China, Hefei 230088, China}

                                                                 \author{Hui Wang}
    \affiliation{Hefei National Laboratory for Physical Sciences at Microscale and School of Physical Sciences,New Cornerstone Science Laboratory, University of Science and Technology of China, Hefei, Anhui, 230026, China}
    \affiliation{CAS Centre for Excellence and Synergetic Innovation Centre in Quantum Information and Quantum Physics, University of Science and Technology of China, Shanghai, 201315, China}
    \affiliation{Hefei National Laboratory, University of Science and Technology of China, Hefei 230088, China}

                                                                 \author{Han-Sen Zhong}
    \affiliation{Shanghai Artificial Intelligence Laboratory, Shanghai 200030, China}
    \affiliation{Shanghai Innovation Institute, Shanghai 200231, China}

                                                                     \author{Ming-Cheng Chen}
    \affiliation{Hefei National Laboratory for Physical Sciences at Microscale and School of Physical Sciences,New Cornerstone Science Laboratory, University of Science and Technology of China, Hefei, Anhui, 230026, China}
    \affiliation{CAS Centre for Excellence and Synergetic Innovation Centre in Quantum Information and Quantum Physics, University of Science and Technology of China, Shanghai, 201315, China}
    \affiliation{Hefei National Laboratory, University of Science and Technology of China, Hefei 230088, China}

                                                                 \author{Xiao Jiang}
    \affiliation{Hefei National Laboratory for Physical Sciences at Microscale and School of Physical Sciences,New Cornerstone Science Laboratory, University of Science and Technology of China, Hefei, Anhui, 230026, China}
    \affiliation{CAS Centre for Excellence and Synergetic Innovation Centre in Quantum Information and Quantum Physics, University of Science and Technology of China, Shanghai, 201315, China}
    \affiliation{Hefei National Laboratory, University of Science and Technology of China, Hefei 230088, China}

                                                                 \author{Li Li}
    \affiliation{Hefei National Laboratory for Physical Sciences at Microscale and School of Physical Sciences,New Cornerstone Science Laboratory, University of Science and Technology of China, Hefei, Anhui, 230026, China}
    \affiliation{CAS Centre for Excellence and Synergetic Innovation Centre in Quantum Information and Quantum Physics, University of Science and Technology of China, Shanghai, 201315, China}
    \affiliation{Hefei National Laboratory, University of Science and Technology of China, Hefei 230088, China}

                                                                 \author{Nai-Le Liu}
    \affiliation{Hefei National Laboratory for Physical Sciences at Microscale and School of Physical Sciences,New Cornerstone Science Laboratory, University of Science and Technology of China, Hefei, Anhui, 230026, China}
    \affiliation{CAS Centre for Excellence and Synergetic Innovation Centre in Quantum Information and Quantum Physics, University of Science and Technology of China, Shanghai, 201315, China}
    \affiliation{Hefei National Laboratory, University of Science and Technology of China, Hefei 230088, China}

    \author{Yu-Hao Deng}
    \affiliation{Hefei National Laboratory for Physical Sciences at Microscale and School of Physical Sciences,New Cornerstone Science Laboratory, University of Science and Technology of China, Hefei, Anhui, 230026, China}
    \affiliation{CAS Centre for Excellence and Synergetic Innovation Centre in Quantum Information and Quantum Physics, University of Science and Technology of China, Shanghai, 201315, China}
    \affiliation{Hefei National Laboratory, University of Science and Technology of China, Hefei 230088, China}

                                \author{Xiao-Long Su}
                            \affiliation{State Key Laboratory of Quantum Optics Technologies and Devices, Institute of Opto-Electronics, Shanxi University, Taiyuan 030006, China}
                            \affiliation{Collaborative Innovation Center of Extreme Optics, Shanxi University, Taiyuan 030006, China}

                                                             \author{Qiang Zhang}
    \affiliation{Hefei National Laboratory for Physical Sciences at Microscale and School of Physical Sciences,New Cornerstone Science Laboratory, University of Science and Technology of China, Hefei, Anhui, 230026, China}
    \affiliation{CAS Centre for Excellence and Synergetic Innovation Centre in Quantum Information and Quantum Physics, University of Science and Technology of China, Shanghai, 201315, China}
    \affiliation{Hefei National Laboratory, University of Science and Technology of China, Hefei 230088, China}
                                \affiliation{Jinan Institute of Quantum Technology and Hefei National Laboratory Jinan Branch, Jinan 250101, China}

                             \author{Chao-Yang Lu}
                             \email{cylu@ustc.edu.cn}
    \affiliation{Hefei National Laboratory for Physical Sciences at Microscale and School of Physical Sciences,New Cornerstone Science Laboratory, University of Science and Technology of China, Hefei, Anhui, 230026, China}
    \affiliation{CAS Centre for Excellence and Synergetic Innovation Centre in Quantum Information and Quantum Physics, University of Science and Technology of China, Shanghai, 201315, China}
    \affiliation{Hefei National Laboratory, University of Science and Technology of China, Hefei 230088, China}

                                 \author{Jian-Wei Pan}
                                 \email{pan@ustc.edu.cn}
    \affiliation{Hefei National Laboratory for Physical Sciences at Microscale and School of Physical Sciences,New Cornerstone Science Laboratory, University of Science and Technology of China, Hefei, Anhui, 230026, China}
    \affiliation{CAS Centre for Excellence and Synergetic Innovation Centre in Quantum Information and Quantum Physics, University of Science and Technology of China, Shanghai, 201315, China}
    \affiliation{Hefei National Laboratory, University of Science and Technology of China, Hefei 230088, China}

	\begin{abstract}

        The creation of large-scale, high-fidelity quantum computers is not only a fundamental scientific endeavour in itself, but also provides  increasingly robust proofs of quantum computational advantage (QCA) in the presence of unavoidable noise and the dynamic competition with classical algorithm improvements. To overcome the biggest challenge of photon-based QCA experiments, photon loss, we report new Gaussian boson sampling (GBS) experiments with 1024 high-efficiency squeezed states injected into a hybrid spatial-temporal encoded, 8176-mode, programmable photonic quantum processor, \textit{Jiuzhang} 4.0, which produces up to 3050 photon detection events. Our experimental results outperform all classical spoofing algorithms, particularly the matrix product state (MPS) method, which was recently proposed to utilise photon loss to reduce the classical simulation complexity of GBS. Using the state-of-the-art MPS algorithm on the most powerful supercomputer EI Capitan, it would take $>10^{42}$ years to construct the required tensor network for simulation, while our \textit{Jiuzhang} 4.0 quantum computer takes  \SI{25.6}{\micro s} to produce a sample. This work establishes a new frontier of QCA and paves the way to fault-tolerant photonic quantum computing hardware.

        \end{abstract}
        
	\maketitle

One of the most interesting aspects of quantum computing is its potential to solve specific problems with exponential speedup compared to classical computers. Recent experimental advancements  \cite{arute2019quantum,zhong2020quantum,PhysRevLett.127.180501,zhong2021phase,madsen2022quantum,deng2023gaussian,morvan2024phase,gaoEstablishingNewBenchmark2025,abanin2025constructiveinterferenceedgequantum} in controlling high-fidelity, large-scale quantum systems have provided growing evidence of quantum computational advantage (QCA) \cite{aaronson2010computationalcomplexitylinearoptics, bouland2019complexity}.
These QCA experiments have, in turn, sparked an intense competition between quantum hardware and classical simulation algorithms 
\cite{panSimulationQuantumCircuits2022,panSolvingSamplingProblem2022a,fuSurpassingSycamoreAchieving2024,panEfficientQuantumCircuit2024,zhaoLeapfroggingSycamoreHarnessing2025,PhysRevLett.124.100502,villalonga2021efficient,quesada2022quadratic,bulmer2022boundary,oh2022classical,oh2023spoofing,oh2024classical}.
For instance, the seminal random circuit sampling experiment \cite{arute2019quantum} was efficiently simulated on GPUs, requiring just 2.9$\%$ of the original computation time and 6.7$\%$  of the power consumption of Sycamore processor \cite{panSolvingSamplingProblem2022a,zhaoLeapfroggingSycamoreHarnessing2025}. This quantum-classical competition continually redefines the boundaries of QCA and drives the development of higher-fidelity, larger-scale quantum computers—a fundamental endeavor.

\begin{figure*}[!htp]
    \centering
    \includegraphics[width = 1.0\linewidth]{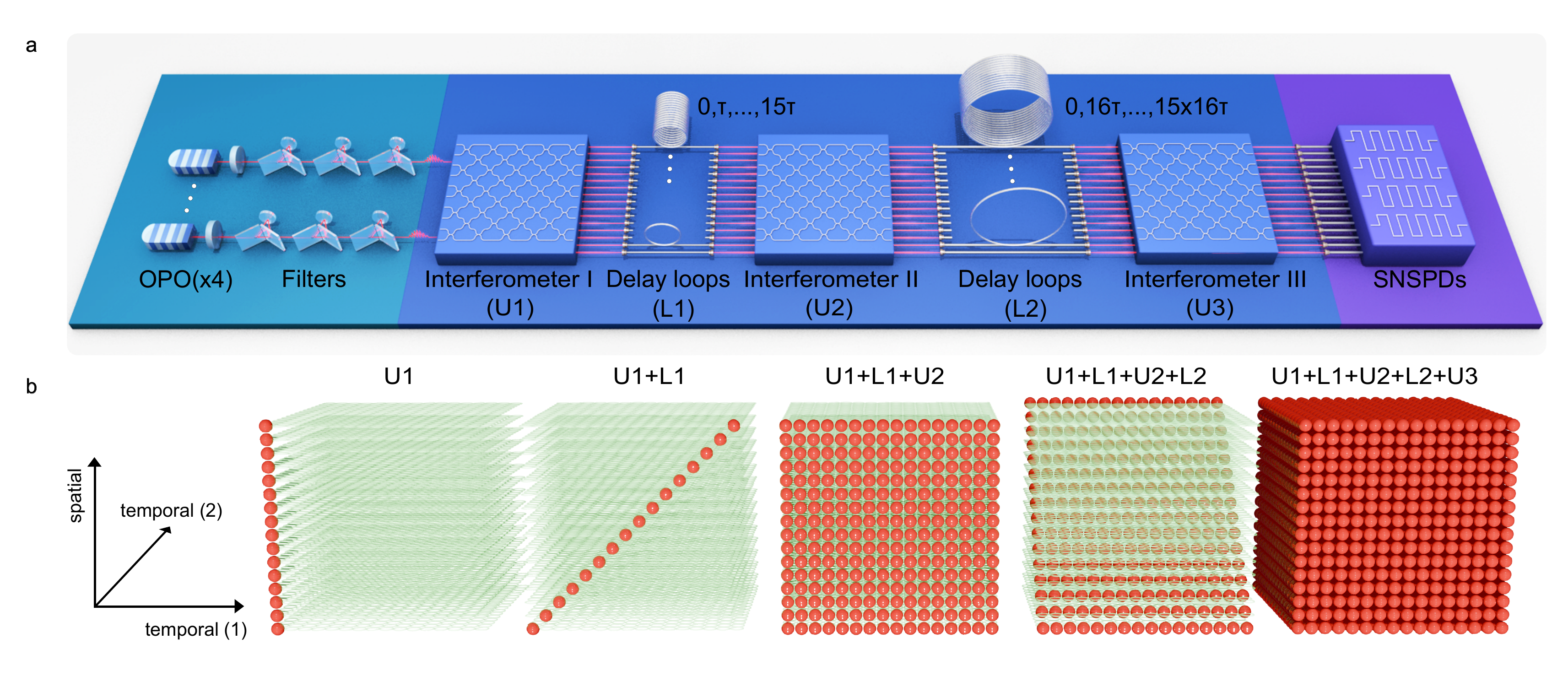}
        \caption{\textbf{a}, The experimental set-up. Four optical parametric oscillators (OPOs) generate the single-mode squeezed states which are filtered and sent into a spatial-temporal hybrid encoding circuit. The circuit is comprised of three 16-mode interferometers connected by two fiber-delay-loop arrays. The first shorter delay loop array is in the range: $[t,t+\tau, …,t+15\tau]$, and the second one is in the range: $[t,t+16\tau, …,t+15\times16\tau]$. \textbf{b}, Evolution of photon population. When photons go through the first interferometer U1 and the subsequent short loop array L1, they are split into 16 sub-beams and appear at the next 16 time bins at $[t, t+\tau,..., t+15\tau]$. Then the 16 sub-beams enter the second interferometer U2 and the subsequent long loop array L2, after which the $i$th loop guides the interior sub-beams to further reach the time bins $[t+i\times16\tau, t+(1+i\times16)\tau,..., t+(15+i\times16)\tau]$. Thus, each input photon can be densely coupled to the next $16^2$ time bins, and spread to all the 16 spatial modes again at the last interferometer U3. A cubic scaling of connectivity ($16^3=4096$) is realized, with one dimension on the spatial domain and two dimensions on the temporal domain. 
}
    \label{fig:1}
\end{figure*}

Gaussian boson sampling (GBS) \cite{hamilton2017gaussian,quesada2018gaussian} is a powerful framework for demonstrating QCA using linear optics, with applications in various practical domains \cite{arrazolaQuantumApproximateOptimization2018,arrazolaUsingGaussianBoson2018,banchiMolecularDockingGaussian,huhBosonSamplingMolecular2015,jahangiriQuantumAlgorithmSimulating2020,jahangiriQuantumAlgorithmSimulating2021,dengSolvingGraphProblems2023a,shangBosonSamplingEnhanced2024a,cimini2025largescalequantumreservoircomputing,gong2025enhancedimagerecognitionusing} and as a building block for universal quantum computing \cite{PhysRevA.101.032315,larsen2025integrated}. 
Prior GBS experiments detected 76, 113, 219, and 255 photon events \cite{zhong2020quantum,zhong2021phase,madsen2022quantum,deng2023gaussian}, with outcomes validated against contemporary classical spoofing algorithms \cite{villalonga2021efficient, oh2023spoofing,PhysRevLett.124.100502,quesada2022quadratic,bulmer2022boundary,oh2022classical}.
However, a primary challenge in all the optical quantum computing experiments is photon loss. For example, previous GBS experiments achieved overall efficiencies of 0.54 with spatial encoding \cite{zhong2021phase} and 0.32 with time-bin encoding \cite{madsen2022quantum}. 

It has long been hypothesized that photon loss could reduce the computational complexity of GBS \cite{ChineseBosonSamplingExperiment2020}. But until very recently, a matrix product state (MPS) based algorithm \cite{oh2024classical} quantitatively evaluated the impact of photon loss in the GBS. The algorithm decomposes the output covariance matrix of lossy GBS into a quantum component, defined by an effective squeezed photon number, and a computationally tractable classical component. Consequently, photon loss diminishes the effective squeezed photon number, thereby reducing GBS computational complexity.

We report a new photonic quantum processor, \textit{Jiuzhang} 4.0 that implements programmable boson sampling with up to 1024 input squeezed states and 8176 output qumodes. 
A schematic drawing of the set-up is shown in Fig. 1a. It consists of three main parts: squeezed light sources, a programmable spatial-temporal hybrid encoding circuit, and a single-photon detection and coincidence system.

Single-mode squeezed states (SMSSs) are created from four optical parametric oscillators (OPOs), which are pumped by 1.6 ns-duration laser pulses centered at 775.7 nm (Fig. S1-S4). To filter out non-degenerate spectral modes, we use three cascaded unbalanced Mach-Zehnder interferometers, which features a near-unity transmission of 99.8$\%$ and \SI{>40}{dB} extinction ratio (Fig. S5). With such a design, the system efficiency of our squeezed light source reaches 92$\%$, significantly higher than the 70$\%$ reported in \cite{madsen2022quantum}. 
 
 With varied pump strength, different squeezing parameters ranging from 0.9 to 1.8 is generated to perform the experiments. 
The experimentally calibrated photon indistinguishability shows a theoretically expected dependence as a function of the squeezing parameter $r$ (Fig. S6-S8), reaching 97$\%$ at $r=1.8$. The ground-truth is modeled as a lossy and partial distinguishable GBS \cite{deng2023gaussian}, which enables better agreement between the experimental observation and the theoretical model (see SM text).

The SMSSs are then sent into a spatial-temporal hybrid encoding circuit, which is designed to achieve high
connectivity, scalability and programmability, simultaneously. The circuit consists of three cascaded fully-connected 16-mode interferometers which are inter-connected by 2 delay loop arrays. The first shorter delay loop array ranges from $[t,t+\tau, …,t+15\tau]$ and the second longer array ranges from $[t,t+16\tau, …,t+15\times16\tau]$. We set the temporal interval $\tau$=\SI{50}{ns} to match the recovery time of the single photon detection channels.

Fig. 1b traces evolution of the photon population inside the circuit. Our cascaded architecture harnesses both spatial and temporal interference, so that information encoded in one degree of freedom can be transformed with the other as the photons propagate. Step 1: After U1 the photons are delocalized over all 16 sptial modes.
Step 2: The subsequent array of delay loops L1 causes the photons to populate the nearest 16 time bins. A second interferometer U2 then redistribute the photons across all $16\times16$ spatial-temporal modes.
Step 3: The second array of delay loops L2 opens a new temporal dimension by delaying the photons with every 16 time bins, and lastly, a third interferometer U3 again mixed all the $16\times16\times16$ spatial-temporal modes.

The connectivity of our circuit grows cubically, while the physical resources, i.e. interferometer size, squeezed light sources, delay lines, and detectors, grow only linearly. This favourable scaling low enables unprecedented ultra-large-scale Gaussian boson sampling experiments involving thousands of highly connected input and output qumodes.


Our photonic processor offers extensive programmability at both the input-state preparation and the interferometric stages. An electro-optic modulators (EOM) acting on the seed laser,  together with the acoustic-optic modulator (AOM) that gates the pumping laser, allow us to prepare arbitrary sequences of single-mode squeezed states as the experimental input. Downstream along the processor, three 16-mode interferometers incorporate thermal tunability, enabling reconfigurability over the spacial modes. Coupling between temporal modes is governed by fiber delay loops whose phase can be programmed with fast piezo-electric stretchers. Taken together, these elements give us control capabilities over all components of our photonic circuits shown in Fig.1a. Once configured at the start of an experiment, the circuit is fixed by active phase-locking loops, ensuring the machine continuously produce samples from the target ground truth distribution.

The output photons are registered by 16 superconducting nanowire single-photon detectors, with an average detection efficiency of 93$\%$ and recovery time of \SI{43}{ns}. The overall system efficiency, including the detection, is measured to be
51$\%$. The phase of the GBS set-up, from the generation of the SMSSs to the interferometers, is stablized to $\sim\lambda/200$ at \SI{1550}{nm} (Fig. S16).

To characterize the performance of our quantum processor, multiple groups of data are collected under different circuit depth and input scale (Fig. S17): the S64 group with 64 input SMSSs and 4336 output qumodes; the M256 group with 256 input SMSSs and 5104 output qumodes; and the L1024 group with 1024 input SMSSs and 8176 output qumodes. The three groups differ only in input scale while share the same circuit. 

Photon number distribution of the L1024 group is plotted in Fig. 2a, we observe up to 3050 photons' coincidence event in the largest experiment, which significantly exceeds all previous results with more than one magnitude. 

\begin{figure}[!htp] 
	\centering  

		\includegraphics[width=0.5\linewidth, width=8.6cm]{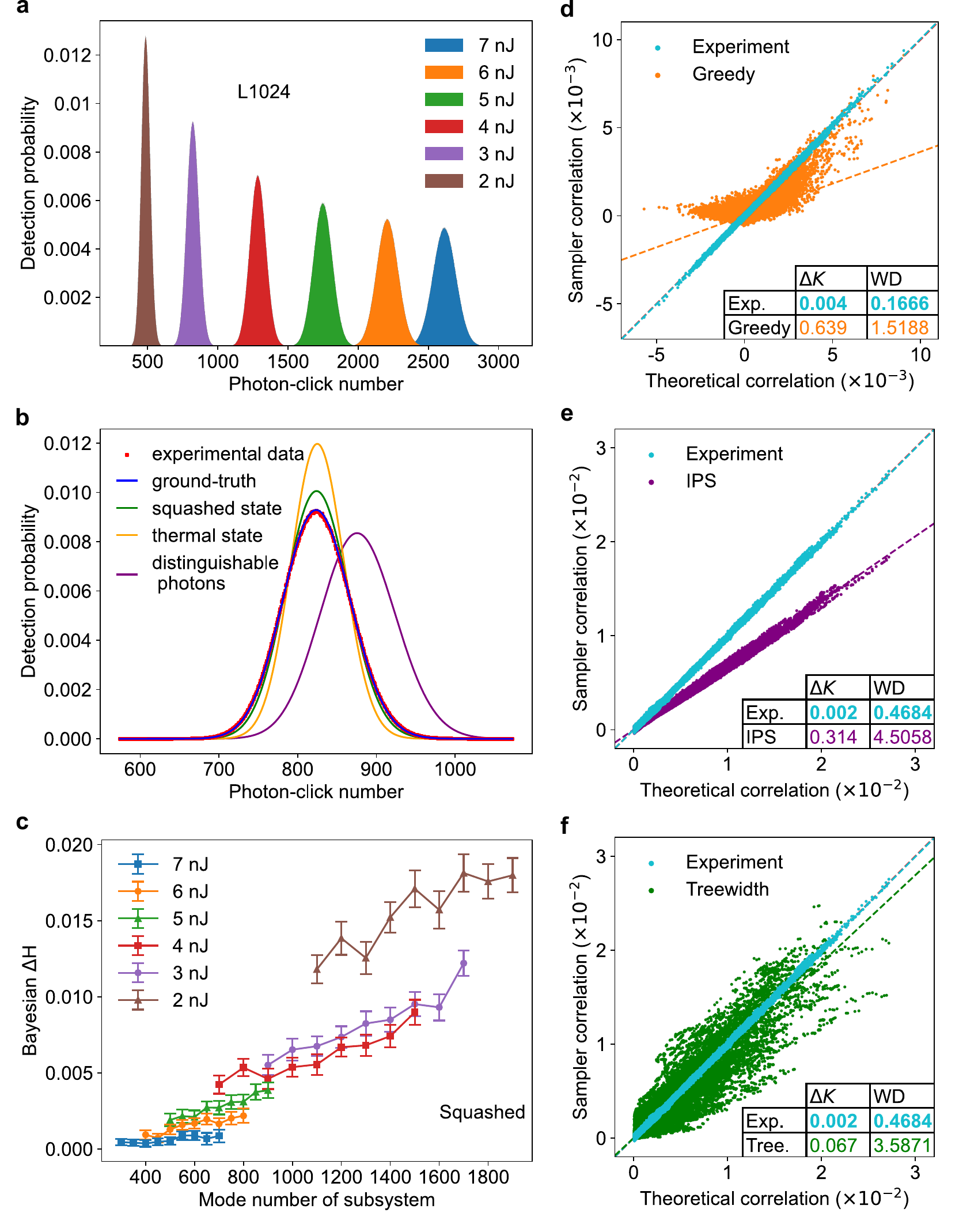}
  
	\caption{\textbf{a}, Obtained photon-click number distribution for the L1024 group at different single-pulse energy. \textbf{b}, A comparison of the photon number distribution of the experimental results, the ground-truth theory and classical mockups, including the squashed state, the thermal state, distinguishable photons. \textbf{c}, Bayesian tests against the squashed state, which are computed under certain photon-click number in terms of the size of considered subsystem. As the size of subsystem grows, the resultant Bayesian score shows a clear trend of increasing which indicates higher confidence for the full system even though it can not be directly evaluated. The error bars are the standard deviation obtained by 100 bootstrapping resamples. \textbf{d}, Scatter plot of the three-order correlation function of the greedy sampler (two-order) in comparison with the experimental results.  \textbf{e},\textbf{f}, Scatter plot of the two-order correlation function of the IPS sampler \textbf{e} and the treewidth sampler \textbf{f} in comparison with the experimental results. The data is from the L1024 group under the highest pump strength at 7 nJ. 10 million samples are used in \textbf{d},\textbf{e},\textbf{f} to evaluate the results. To count for the statistical fluctuation due to finite amount of samples, we compute the two-norm distance of the correlation function by weighting on the modulus of the data points’ theoretical value (WD denotes weighted distance). Our experimental sampler agrees well with the ground-truth and significantly outperforms all these mockup samplers in terms of the correlation function.}
	\label{level}
\end{figure}

 Firstly, we compare the experimental photon number distribution with the theoretical prediction of the groundtruth and the mockups, which conveys global information of the ensemble. As shown in Fig. 2b, the experimental distribution overlaps well with the groundtruth while strongly deviates from all the classical mockups.

Then we continue with the Bayesian test, a robust statistical tool to distinguish the theoretical model of adversary hypotheses from the groundtruth based on the likelihood to generate the experimental samples. The Bayesian test score $\Delta H$ is defined as the difference between the ground-truth model $H_0$ and the classical adversary model $H_1$ on a set of $n$-photon-click samples \cite{deng2023gaussian}. A Bayesian test score $\Delta H >0$ means that the experimental samples are more likely from the groundtruth GBS rather than the mockup and a higher score indicates higher confidence.

We test the classical mockup states which can maximally approximate the experiment’s quantum light sources under photon
loss: the squashed state \cite{madsen2022quantum, deng2023gaussian}. 
We start from a subsystem with fewer output bosonic modes,
and gradually increase the subsystem size \cite{zhong2021phase}. The results in Fig. 2c show that not only all the Bayesian scores are higher than zero which indicates that the experimental samples are more likely generated from the ground-truth distribution, there is also a clear trend of increasing as the size of subsystem ramps up, because
more complete information of the entire system is included in the test. From such an observation we can conclusively infer that stronger Bayesian confidence will be expected for the full system even though it can not be directly evaluated.

Another important tool to validate the experimental GBS is the correlation function benchmarking \cite{zhong2021phase,deng2023gaussian}. We use the slope $K$ of linear fit between the theoretical correlation and the counted sampler correlation to quantify the performance of various of samplers. Here we define $\Delta K=|K_{\textrm{sampler}}-1|$ as the metric of deviation from the ground-truth. Smaller $\Delta K$ indicates better performance of the sampler. For the ideal GBS, $\Delta K = 0$.

Fig. 2d shows a comparison of the third-order correlation between our experimental samples and the mockup samples from the so-called greedy sampler proposed in Ref. \cite{villalonga2021efficient}, which aims to spoof GBS samples by reproducing the low-order marginals. The mockup is shown not able to capture higher-order correlations, and thus is conclusively ruled out. Similarly, Fig. 2e rules out the independent pairs and singles (IPS) sampler 
designed in \cite{bulmer2022boundary} to emulate experiments of limited quantum
interference. 

The treewidth sampler proposed in \cite{oh2022classical} uses local connectivity of GBS circuits to perform approximate simulation, which has computational complexity exponentially dependent on the chosen treewidth parameter. For the largest experiment, we set the treewidth to 801 (a regime beyond practical implementation) to check its approximation performance in terms of the correlation function.  Fig. 2f shows a significant deviation from the ground-truth, which indicates that this algorithm is not able to simulate our experiments.

\begin{figure*}[!htp]
    \centering
    \includegraphics[width = 1\linewidth]{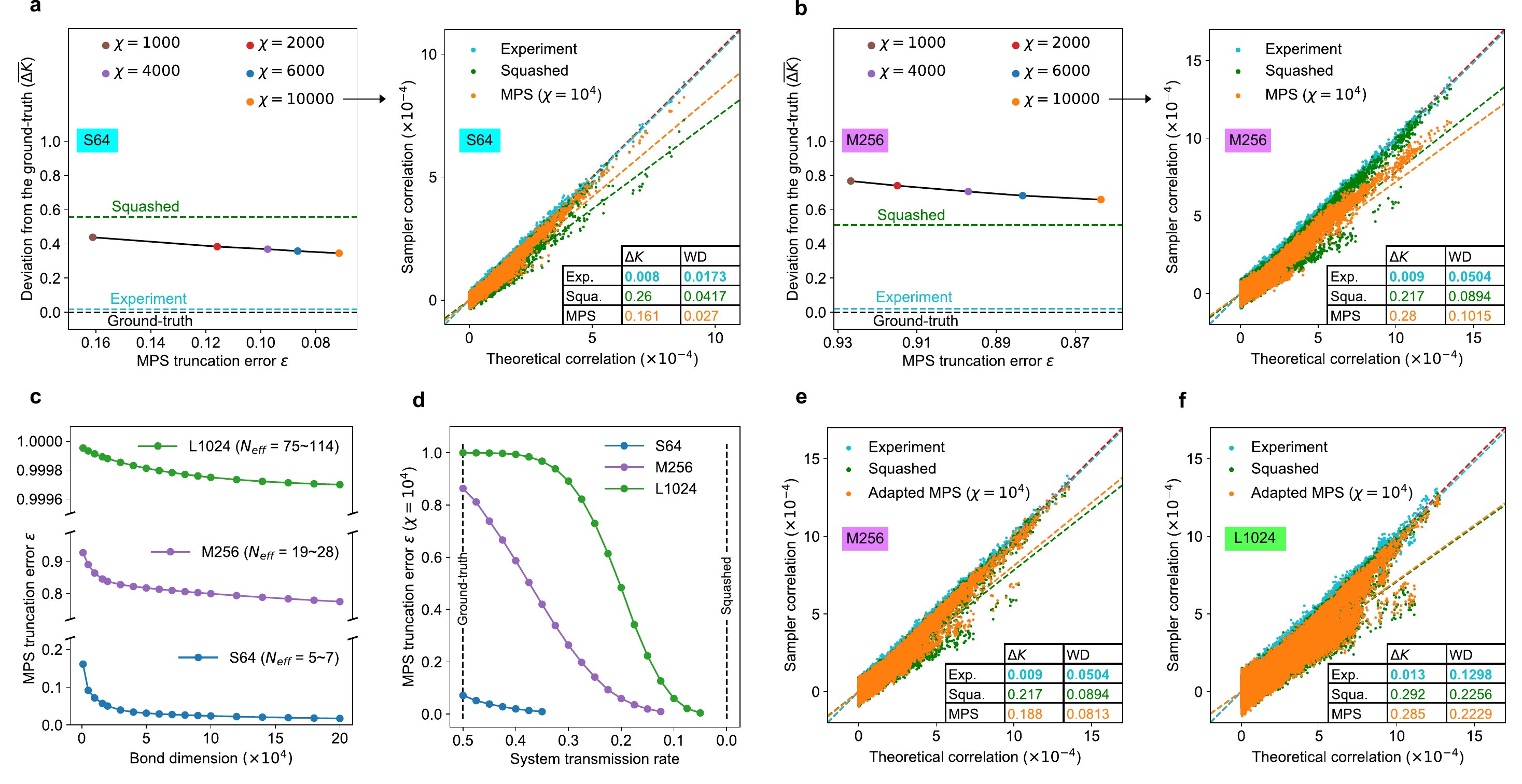}
    \caption{ \textbf{a},\textbf{b} MPS truncation error and $\Delta K$ of the two-order correlation function counted from MPS samples, for the S64 experiment \textbf{a} and the M256 experiment \textbf{b}, under progressively increasing bond dimension $\chi$. 
    When presenting the left panels in \textbf{a} and \textbf{b}, we define the normalized deviation quantity as $\overline{\Delta K} = \Delta K/\Delta K_0$, where $\Delta K_0$ is the $\Delta K$ obtained with MPS under $\chi=0$, i.e, ignoring the quantum part $V_p$ but only remaining the thermal displacement in $W$. 
    \textbf{c}, Truncation error of MPS in approximating the ground-truth of the S64, M256 and L1024 experiments, in terms of the chosen bond dimension up to $\chi$=$2\times10^5$. The effective squeezed photon number $N_\textrm{eff}$ of the three groups is also labeled respectively.
    \textbf{d}, With a fixed bond dimension $\chi$=$10^4$, the truncation error’s dependence on the scaling down of transmission (keeping the mean photon number fixed) for the three groups of experiments. 
    The ground-truth is labeled as the left dash line at the start point, and the squashed state corresponds to the asymptotic limit under excess loss which is labeled as the right dash line. 
    \textbf{e},\textbf{f}, Scatter plot of the two-order correlation function counted from the experimental samples, squashed state samples, and the adapted MPS ($\chi$=$10^4$) samples, for the M256 experiment \textbf{e} and the L1024 experiment \textbf{f}. 
    5.5 million samples are used in \textbf{a},\textbf{b}, and \textbf{e}. 10 million samples are used in \textbf{f}.}
    \label{fig:3}
\end{figure*}

Lastly, we demonstrate that our experiment vastly exceeds the reach of the MPS algorithm proposed in Ref.\cite{oh2024classical}, which is currently the most powerful classical algorithm to simulate lossy GBS.
The MPS algorithm works by first decomposing the output covariance matrix $V$ of a lossy GBS device into a quantum part $V_p$ and a classical part $W$, $V=V_p+W$, where $V_p$ represents an ideal but smaller-scale GBS state, and $W$ represents random displacement and can be efficiently simulated. The core of the algorithm is to construct a MPS in the Fock basis to approximately simulate the evolution of the quantum part $V_p$. Because photon loss can reduce the effective squeezing, $V_p$ typically contains fewer photons than the physical device, allowing the MPS algorithm to simulate lossy GBS more efficiently than exact approaches.

 Two crucial parameters determine the algorithm's accuracy and computational complexity: (1), The bond dimension $\chi$ which controls the truncation error when approximating $V_p$ with the MPS. Both runtime and memory cost scale as $ O(\chi^2)$. And (2), The effective squeezed photon number $N_{\textrm{eff}}$ which accounts for the hafnian calculation cost during tensor construction and evolution. The runtime grows exponentially on $N_{\textrm{eff}}$.

To exclude an MPS simulation of our experiment, we show that no $\chi$ and $N_{\textrm{eff}}$ compatible with present-day resources can reproduce the statistics of our device within experimental uncertainty. Our benchmarking follows two complementary approaches: 

\textit{Targeting ideal ground truth distribution.} By fixing the target of MPS algorithm to be the ideal ground truth, we evaluate the truncation error and output statistics error of MPS simulator with increasing experimental scale. We show that for our largest-scale experiment, the computational resources significantly surpass the capabilities of supercomputers, far before the MPS error can be brought down to the level achieved experimentally. 

\textit{Targeting modified distribution approximating the ground truth with a tractable $N_{\textrm{eff}}$.} As an alternative investigation, we target the MPS algorithm to approximate the ground-truth wtih less effective photon number, keeping the truncation error sufficiently low and the runtime manageable. We find the sampler's output statistics deviate sharply from the ideal distribution and fail standard validation tests.

We begin detailing our first approach by investigating truncation error's growth at various experimental scale, and consequently how the statistics of MPS simulator deteriorate.
A useful reference is the squashed-state model, obtained by replacing the quantum part $V_p$ with vacuum, which represents the classical-state simulator closest to our ground-truth, i.e. a baseline from which MPS must improve.

For the S64 data set, we draw 5.5 million MPS samples for several bond dimensions and benchmark them with the two-order correlation metric $\Delta K$ \cite{deng2023gaussian, oh2024classical}. On the left panel of Fig. 3a we show the normalized $\overline{\Delta K}$ versus the truncation error $\varepsilon$, and on the right panel we plot the detailed results for the case of $\chi=10^4$. As $\chi$ increases from $10^3$ to $10^4$, the $\varepsilon$ and the simulated $\overline{\Delta K}$ is improved from 0.16 to 0.07 and from 0.44 to 0.34, respectively, converging towards the ground truth ($\overline{\Delta K}=0$). For this data set, the MPS outperforms the squashed state baseline ($\overline{\Delta K}=0.56$), however, still significantly deviates from the experiment results ($\overline{\Delta K}=0.02$) even for the largest $\chi$.

The M256 experiment exhibits a different picture, as shown in Fig. 3b. Even at the largest bond dimension of $10^4$, the truncation error reaches 0.86. More importantly, the two-order correlation metric $\overline{\Delta K}$ grows to 0.66, a point that is worse than the squashed-state baseline of 0.51. This is because, at this error level, the MPS method loses so much information that the simulated distribution statistics drifts already far from the ground truth.

The experimental size-dependence of the MPS simualtor is already evident from the observations above. We further present in Fig. 3c the evaluated truncation error for all the three data sets up to $\chi = 2 \times 10^5$.
For the L1024 data set, the truncation error $\varepsilon$ reaches 0.9999 under $\chi = 10^4$, a level at which the MPS-simulated statistics would be essentially meaningless. Yet, even reaching this poor accuracy is already beyond classical computing. For the L1024 data set, $N_{\textrm{eff}}$ reaches up to 113.5. Calculating a single MPS element requires exponential time cost hafnian evaluation. For example, running the hardest L1024 sample with $\chi=10^4$ would require more than $2\times 10^6$ years on a $10^5$-GPU cluster such as xAI-Colossus \cite{xai}. Consequently, no realistic classical computing resources can bring the MPS algorithm anywhere near the accuracy achieved by our experiment.

We next investigate the complementary aspect that whether a simulatable MPS sampler, with smaller $N_{\textrm{eff}}$ and therefore manageable run time as well as negligible truncation error, could approximate the ideal ground truth closer than our experiment does. Concretely, we target the MPS algorithm to simulate a distribution that is interpolated between the ideal ground truth and the squashed state, with the latter being the classical state closest to the ground truth. In order to construct such a sampler, we artificially reduce the circuit transmission to lower $N_{\textrm{eff}}$ while increasing the input squeezing so that the mean photon number left unchanged.

With the bond dimension fixed to $10^4$ we scan and choose, for each experimental size, a transmission that yields a favorable truncation error of 0.01. As shown in Fig. 3d, the larger the experimental scale, the more the algorithm reverts towards the squashed state, which is exactly as expected. Running this specially tailored MPS simulator on the M256 and L1024 data set (Fig. 3e-f) does improve upon the squashed state baseline, yet its outputs still differ markedly from the ground truth in comparison with our experimental distribution.

These results exclude all possibility that the MPS algorithm could reproduce the ground truth distribution better than our experiment. We note that the observations here are in consistent with the asymptotic criterion of Ref. \cite{oh2024classical}, which states that the MPS algorithm would keep being efficient as long as the transmission $\eta$ decays as $\eta=o(1/\sqrt{K})$, where $K$ is the number of input squeezed states. In our experimental architecture the scale-up is achieved at a fixed transmission rate, in which case the MPS is driven outside its efficient regime as the experimental size grows.

\begin{figure*}[!htp]
    \centering
    \includegraphics[width = 0.85\linewidth]{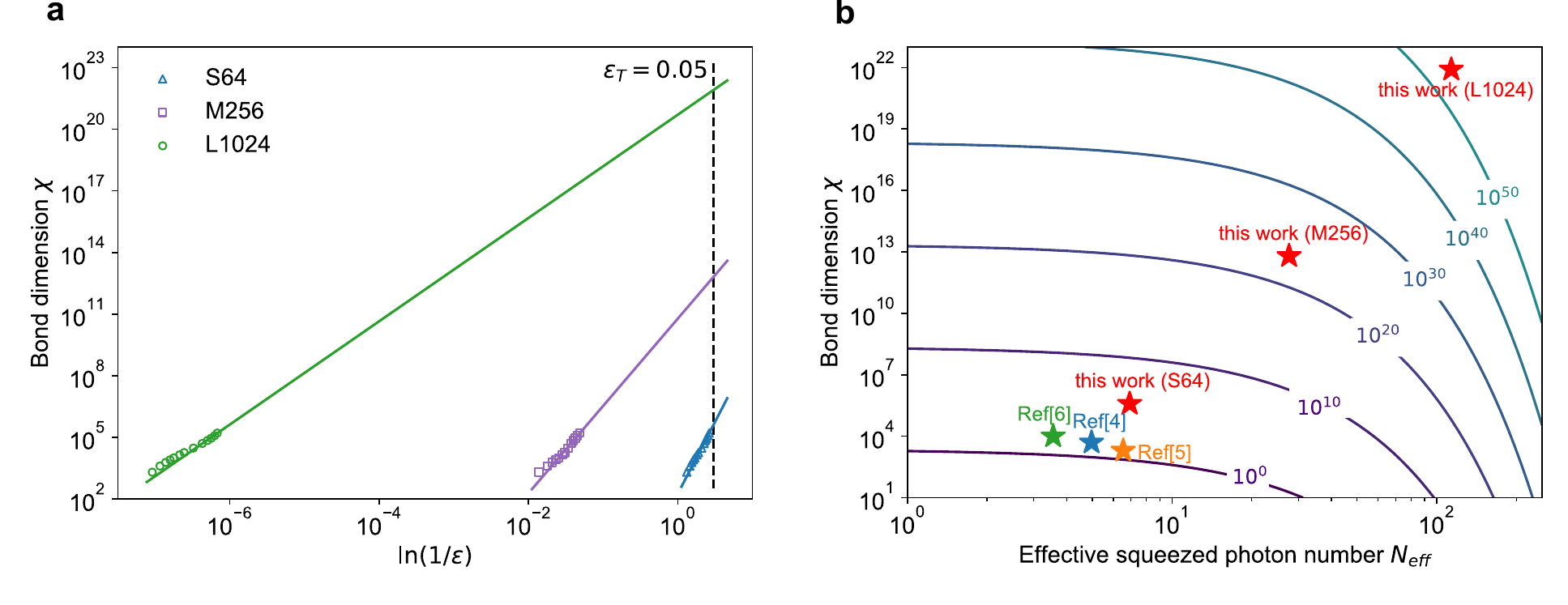}
    \caption{\textbf{Estimation of the quantum speedup against MPS simulation.}
 \textbf{a}, Fitted relation between the bond dimension $\chi$ and the truncation error $\varepsilon$, and the extrapolation to a target truncation error $\varepsilon_T$ to estimate the required size of $\chi$. 
 \textbf{b}, Estimated speedup ratio against MPS simulation, which is based on EI Capitan, currently the most powerful supercomputer. The results are presented in terms of the two crucial factors, the bond dimension $\chi$ and the effective squeezed photon number $N_\textrm{eff}$. 
 }
    \label{fig:3}
\end{figure*}

Finally, we benchmark the QCA of our experiments by estimating the computational cost it would take for the state-of-the-art classical approach--the MPS algorithm \cite{oh2024classical}--to generate equivalent samples.
For simulating an $M$-mode GBS experiment with a local Fock-space cut-off dimension $d$, the asymptotic runtime of the MPS algorithm is
\begin{equation}
T_\textrm{MPS} = O(Md \chi(\varepsilon)^2 2^{ \frac{N_\textrm{eff}}{2} })
\end{equation}
where $\chi(\varepsilon)$ is the bond dimension at a truncation error $\varepsilon$, and $N_{\textrm{eff}}$ is the effective squeezed photon number. 

The computational cost is dominated by the two exponential factors, $2^{{N_\textrm{eff}}/2}$ and $\chi(\varepsilon)^2$. $N_\textrm{eff}$ is computed exactly from the squeezing parameters and the interferometer matrix (Table S2). 
By contrast, computing $\chi(\varepsilon)$ exactly is infeasible for the QCA regime, because the required tensor calculations exceed any realistic memory budget. 
Instead, we extrapolate from the tractable regime to estimate the required $\chi$. Following Ref. \cite{oh2024classical} which shows that $\chi(\varepsilon) = O(\operatorname{polylog}(1/\varepsilon))$, we find our data to be well described by  $\chi(\varepsilon) = A(\ln(1/\varepsilon))^n$ (Fig. S19), with coefficients $A,n$ determined by least-squares fitting. Choosing a conservative truncation error of $\varepsilon_T=0.05$ yields a lower bound of the required $\chi>8\times10^{21}$ for our largest L1024 data set, as shown in Fig. 4a.

Having determined the least required $\chi$, we then estimate the time cost for the MPS algorithm using Eq. (1). We choose EI Capitan--the currently most powerful supercomputer \cite{November2024TOP500}, as the classical baseline. To obtain the empirical prefactor of Eq.(1), we first benchmark the runtime of the MPS simulation in a regime where direct simulation is feasible \cite{fastMPS} . We then scale the measured runtime to the intractable regime according to Eq. (1), implicitly granting the classical supercomputer unlimited memory—an assumption that favours the classical side. Fig. 4b summarizes the projected runtime advantage ratios for all three experimental instances reported in this work, as well as earlier GBS demonstrations. For our largest experiment, EI Capitan would require more than $10^{42}$ years to produce a sample, while our quantum processor takes only \SI{25.6}{\micro s}. This corresponds to a tremendous quantum speedup ratio exceeding $10^{54}$.

In conclusion, we have realized the largest boson sampling experiment so far, featuring up to 1024 input squeezed states and 8176 output qumodes, yielding a Hilbert space dimension to $\sim 10^{2461}$.
This configuration overcomes the most important noise for photonic quantum computing systems---photon loss---to rule out all classical spoofing models and achieve a robust and overwhelming $10^{54}$ QCA over the state-of-the-art classical algorithms. The development of low-loss squeezed light sources and programmable spatial-temporal hybrid encoding circuits not only immediately allows us to control 3D massive highly-entangled qumodes cluster states in the near future, but also paves the way towards the next generation of fault-tolerant photonic quantum computing hardware .

\textbf{data availability}:
The dataset of this experiment can be accessed via \cite{rawData}.
\textbf{acknowledgements}:
We thank Changhun Oh, Xiang Zhang, Junhua Zhang, and Haowen Cheng for helpful discussions and assistance.
\textbf{funding}: This work was supported by the National Natural
Science Foundation of China (No. 12434015), the National Key RD
Program of China (No. 2019YFA0308700), the Chinese
Academy of Sciences, the Anhui Initiative in Quantum
Information Technologies, and the Science and
Technology Commission of Shanghai Municipality
(No. 2019SHZDZX01). Innovation Program for
Quantum Science and Technology (No. ZD0202010000, No. 2024ZD0302403),
and the New Cornerstone Science Foundation. \textbf{contributions}: C.-Y.L. and J.-W.P. conceived and supervised the project. H.-L.L. and C.-Y.L designed the squeezed light sources and the hybrid circuit. H.-L.L., H.S., S.-Q.G., Y.-C.G., H.-Y.T. and M.-H.J. carried out the optical experiment and collected the data. Q.W.,X.J., Y.-K.S.,D.-Z.W.,X.-L.S., M.-Y.Z.  F.-X.C., L.-B.L. and L.Z. provided key elements. Y.-J.C., L.-S.S., P.-Y.Y. and J.-S.C. performed the MPS simulation and data validation on supercomputers. 
H.-L.L, S.-Q.G, Y.-C.G., H.S., Y.-H.D, C.-Y.L. and J.-W.P. prepared the manuscript with input from all authors.

 \bibliography{main.bbl}

\begin{thebibliography}{42}%
\makeatletter
\providecommand \@ifxundefined [1]{%
 \@ifx{#1\undefined}
}%
\providecommand \@ifnum [1]{%
 \ifnum #1\expandafter \@firstoftwo
 \else \expandafter \@secondoftwo
 \fi
}%
\providecommand \@ifx [1]{%
 \ifx #1\expandafter \@firstoftwo
 \else \expandafter \@secondoftwo
 \fi
}%
\providecommand \natexlab [1]{#1}%
\providecommand \enquote  [1]{``#1''}%
\providecommand \bibnamefont  [1]{#1}%
\providecommand \bibfnamefont [1]{#1}%
\providecommand \citenamefont [1]{#1}%
\providecommand \href@noop [0]{\@secondoftwo}%
\providecommand \href [0]{\begingroup \@sanitize@url \@href}%
\providecommand \@href[1]{\@@startlink{#1}\@@href}%
\providecommand \@@href[1]{\endgroup#1\@@endlink}%
\providecommand \@sanitize@url [0]{\catcode `\\12\catcode `\$12\catcode
  `\&12\catcode `\#12\catcode `\^12\catcode `\_12\catcode `\%12\relax}%
\providecommand \@@startlink[1]{}%
\providecommand \@@endlink[0]{}%
\providecommand \url  [0]{\begingroup\@sanitize@url \@url }%
\providecommand \@url [1]{\endgroup\@href {#1}{\urlprefix }}%
\providecommand \urlprefix  [0]{URL }%
\providecommand \Eprint [0]{\href }%
\providecommand \doibase [0]{https://doi.org/}%
\providecommand \selectlanguage [0]{\@gobble}%
\providecommand \bibinfo  [0]{\@secondoftwo}%
\providecommand \bibfield  [0]{\@secondoftwo}%
\providecommand \translation [1]{[#1]}%
\providecommand \BibitemOpen [0]{}%
\providecommand \bibitemStop [0]{}%
\providecommand \bibitemNoStop [0]{.\EOS\space}%
\providecommand \EOS [0]{\spacefactor3000\relax}%
\providecommand \BibitemShut  [1]{\csname bibitem#1\endcsname}%
\let\auto@bib@innerbib\@empty
\bibitem [{\citenamefont {Arute}\ \emph {et~al.}(2019)\citenamefont {Arute},
  \citenamefont {Arya}, \citenamefont {Babbush}, \citenamefont {Bacon},
  \citenamefont {Bardin}, \citenamefont {Barends}, \citenamefont {Biswas},
  \citenamefont {Boixo}, \citenamefont {Brandao}, \citenamefont {Buell} \emph
  {et~al.}}]{arute2019quantum}%
  \BibitemOpen
  \bibfield  {author} {\bibinfo {author} {\bibfnamefont {F.}~\bibnamefont
  {Arute}}, \bibinfo {author} {\bibfnamefont {K.}~\bibnamefont {Arya}},
  \bibinfo {author} {\bibfnamefont {R.}~\bibnamefont {Babbush}}, \bibinfo
  {author} {\bibfnamefont {D.}~\bibnamefont {Bacon}}, \bibinfo {author}
  {\bibfnamefont {J.~C.}\ \bibnamefont {Bardin}}, \bibinfo {author}
  {\bibfnamefont {R.}~\bibnamefont {Barends}}, \bibinfo {author} {\bibfnamefont
  {R.}~\bibnamefont {Biswas}}, \bibinfo {author} {\bibfnamefont
  {S.}~\bibnamefont {Boixo}}, \bibinfo {author} {\bibfnamefont {F.~G.}\
  \bibnamefont {Brandao}}, \bibinfo {author} {\bibfnamefont {D.~A.}\
  \bibnamefont {Buell}}, \emph {et~al.},\ }\bibfield  {title} {\bibinfo {title}
  {Quantum supremacy using a programmable superconducting processor},\ }\href
  {https://www.nature.com/articles/s41586-019-1666-5} {\bibfield  {journal}
  {\bibinfo  {journal} {Nature}\ }\textbf {\bibinfo {volume} {574}},\ \bibinfo
  {pages} {505} (\bibinfo {year} {2019})}\BibitemShut {NoStop}%
\bibitem [{\citenamefont {Zhong}\ \emph {et~al.}(2020)\citenamefont {Zhong},
  \citenamefont {Wang}, \citenamefont {Deng}, \citenamefont {Chen},
  \citenamefont {Peng}, \citenamefont {Luo}, \citenamefont {Qin}, \citenamefont
  {Wu}, \citenamefont {Ding}, \citenamefont {Hu} \emph
  {et~al.}}]{zhong2020quantum}%
  \BibitemOpen
  \bibfield  {author} {\bibinfo {author} {\bibfnamefont {H.-S.}\ \bibnamefont
  {Zhong}}, \bibinfo {author} {\bibfnamefont {H.}~\bibnamefont {Wang}},
  \bibinfo {author} {\bibfnamefont {Y.-H.}\ \bibnamefont {Deng}}, \bibinfo
  {author} {\bibfnamefont {M.-C.}\ \bibnamefont {Chen}}, \bibinfo {author}
  {\bibfnamefont {L.-C.}\ \bibnamefont {Peng}}, \bibinfo {author}
  {\bibfnamefont {Y.-H.}\ \bibnamefont {Luo}}, \bibinfo {author} {\bibfnamefont
  {J.}~\bibnamefont {Qin}}, \bibinfo {author} {\bibfnamefont {D.}~\bibnamefont
  {Wu}}, \bibinfo {author} {\bibfnamefont {X.}~\bibnamefont {Ding}}, \bibinfo
  {author} {\bibfnamefont {Y.}~\bibnamefont {Hu}}, \emph {et~al.},\ }\bibfield
  {title} {\bibinfo {title} {Quantum computational advantage using photons},\
  }\href {https://www.science.org/doi/10.1126/science.abe8770} {\bibfield
  {journal} {\bibinfo  {journal} {Science}\ }\textbf {\bibinfo {volume}
  {370}},\ \bibinfo {pages} {1460} (\bibinfo {year} {2020})}\BibitemShut
  {NoStop}%
\bibitem [{\citenamefont {Wu}\ \emph {et~al.}(2021)\citenamefont {Wu},
  \citenamefont {Bao}, \citenamefont {Cao}, \citenamefont {Chen}, \citenamefont
  {Chen}, \citenamefont {Chen}, \citenamefont {Chung}, \citenamefont {Deng},
  \citenamefont {Du} \emph {et~al.}}]{PhysRevLett.127.180501}%
  \BibitemOpen
  \bibfield  {author} {\bibinfo {author} {\bibfnamefont {Y.}~\bibnamefont
  {Wu}}, \bibinfo {author} {\bibfnamefont {W.-S.}\ \bibnamefont {Bao}},
  \bibinfo {author} {\bibfnamefont {S.}~\bibnamefont {Cao}}, \bibinfo {author}
  {\bibfnamefont {F.}~\bibnamefont {Chen}}, \bibinfo {author} {\bibfnamefont
  {M.-C.}\ \bibnamefont {Chen}}, \bibinfo {author} {\bibfnamefont
  {X.}~\bibnamefont {Chen}}, \bibinfo {author} {\bibfnamefont {T.-H.}\
  \bibnamefont {Chung}}, \bibinfo {author} {\bibfnamefont {H.}~\bibnamefont
  {Deng}}, \bibinfo {author} {\bibfnamefont {Y.}~\bibnamefont {Du}}, \emph
  {et~al.},\ }\bibfield  {title} {\bibinfo {title} {Strong quantum
  computational advantage using a superconducting quantum processor},\ }\href
  {https://doi.org/10.1103/PhysRevLett.127.180501} {\bibfield  {journal}
  {\bibinfo  {journal} {Physical Review Letters}\ }\textbf {\bibinfo {volume}
  {127}},\ \bibinfo {pages} {180501} (\bibinfo {year} {2021})}\BibitemShut
  {NoStop}%
\bibitem [{\citenamefont {Zhong}\ \emph {et~al.}(2021)\citenamefont {Zhong},
  \citenamefont {Deng}, \citenamefont {Qin}, \citenamefont {Wang},
  \citenamefont {Chen}, \citenamefont {Peng}, \citenamefont {Luo},
  \citenamefont {Wu}, \citenamefont {Gong}, \citenamefont {Su} \emph
  {et~al.}}]{zhong2021phase}%
  \BibitemOpen
  \bibfield  {author} {\bibinfo {author} {\bibfnamefont {H.-S.}\ \bibnamefont
  {Zhong}}, \bibinfo {author} {\bibfnamefont {Y.-H.}\ \bibnamefont {Deng}},
  \bibinfo {author} {\bibfnamefont {J.}~\bibnamefont {Qin}}, \bibinfo {author}
  {\bibfnamefont {H.}~\bibnamefont {Wang}}, \bibinfo {author} {\bibfnamefont
  {M.-C.}\ \bibnamefont {Chen}}, \bibinfo {author} {\bibfnamefont {L.-C.}\
  \bibnamefont {Peng}}, \bibinfo {author} {\bibfnamefont {Y.-H.}\ \bibnamefont
  {Luo}}, \bibinfo {author} {\bibfnamefont {D.}~\bibnamefont {Wu}}, \bibinfo
  {author} {\bibfnamefont {S.-Q.}\ \bibnamefont {Gong}}, \bibinfo {author}
  {\bibfnamefont {H.}~\bibnamefont {Su}}, \emph {et~al.},\ }\bibfield  {title}
  {\bibinfo {title} {Phase-programmable gaussian boson sampling using
  stimulated squeezed light},\ }\href
  {https://journals.aps.org/prl/abstract/10.1103/PhysRevLett.127.180502}
  {\bibfield  {journal} {\bibinfo  {journal} {Physical review letters}\
  }\textbf {\bibinfo {volume} {127}},\ \bibinfo {pages} {180502} (\bibinfo
  {year} {2021})}\BibitemShut {NoStop}%
\bibitem [{\citenamefont {Madsen}\ \emph {et~al.}(2022)\citenamefont {Madsen},
  \citenamefont {Laudenbach}, \citenamefont {Askarani}, \citenamefont
  {Rortais}, \citenamefont {Vincent}, \citenamefont {Bulmer}, \citenamefont
  {Miatto}, \citenamefont {Neuhaus}, \citenamefont {Helt}, \citenamefont
  {Collins} \emph {et~al.}}]{madsen2022quantum}%
  \BibitemOpen
  \bibfield  {author} {\bibinfo {author} {\bibfnamefont {L.~S.}\ \bibnamefont
  {Madsen}}, \bibinfo {author} {\bibfnamefont {F.}~\bibnamefont {Laudenbach}},
  \bibinfo {author} {\bibfnamefont {M.~F.}\ \bibnamefont {Askarani}}, \bibinfo
  {author} {\bibfnamefont {F.}~\bibnamefont {Rortais}}, \bibinfo {author}
  {\bibfnamefont {T.}~\bibnamefont {Vincent}}, \bibinfo {author} {\bibfnamefont
  {J.~F.}\ \bibnamefont {Bulmer}}, \bibinfo {author} {\bibfnamefont {F.~M.}\
  \bibnamefont {Miatto}}, \bibinfo {author} {\bibfnamefont {L.}~\bibnamefont
  {Neuhaus}}, \bibinfo {author} {\bibfnamefont {L.~G.}\ \bibnamefont {Helt}},
  \bibinfo {author} {\bibfnamefont {M.~J.}\ \bibnamefont {Collins}}, \emph
  {et~al.},\ }\bibfield  {title} {\bibinfo {title} {Quantum computational
  advantage with a programmable photonic processor},\ }\href
  {https://www.nature.com/articles/s41586-022-04725-x} {\bibfield  {journal}
  {\bibinfo  {journal} {Nature}\ }\textbf {\bibinfo {volume} {606}},\ \bibinfo
  {pages} {75} (\bibinfo {year} {2022})}\BibitemShut {NoStop}%
\bibitem [{\citenamefont {Deng}\ \emph
  {et~al.}(2023{\natexlab{a}})\citenamefont {Deng}, \citenamefont {Gu},
  \citenamefont {Liu}, \citenamefont {Gong}, \citenamefont {Su}, \citenamefont
  {Zhang}, \citenamefont {Tang}, \citenamefont {Jia}, \citenamefont {Xu},
  \citenamefont {Chen} \emph {et~al.}}]{deng2023gaussian}%
  \BibitemOpen
  \bibfield  {author} {\bibinfo {author} {\bibfnamefont {Y.-H.}\ \bibnamefont
  {Deng}}, \bibinfo {author} {\bibfnamefont {Y.-C.}\ \bibnamefont {Gu}},
  \bibinfo {author} {\bibfnamefont {H.-L.}\ \bibnamefont {Liu}}, \bibinfo
  {author} {\bibfnamefont {S.-Q.}\ \bibnamefont {Gong}}, \bibinfo {author}
  {\bibfnamefont {H.}~\bibnamefont {Su}}, \bibinfo {author} {\bibfnamefont
  {Z.-J.}\ \bibnamefont {Zhang}}, \bibinfo {author} {\bibfnamefont {H.-Y.}\
  \bibnamefont {Tang}}, \bibinfo {author} {\bibfnamefont {M.-H.}\ \bibnamefont
  {Jia}}, \bibinfo {author} {\bibfnamefont {J.-M.}\ \bibnamefont {Xu}},
  \bibinfo {author} {\bibfnamefont {M.-C.}\ \bibnamefont {Chen}}, \emph
  {et~al.},\ }\bibfield  {title} {\bibinfo {title} {Gaussian boson sampling
  with pseudo-photon-number-resolving detectors and quantum computational
  advantage},\ }\href
  {https://journals.aps.org/prl/abstract/10.1103/PhysRevLett.131.150601}
  {\bibfield  {journal} {\bibinfo  {journal} {Physical review letters}\
  }\textbf {\bibinfo {volume} {131}},\ \bibinfo {pages} {150601} (\bibinfo
  {year} {2023}{\natexlab{a}})}\BibitemShut {NoStop}%
\bibitem [{\citenamefont {Morvan}\ \emph {et~al.}(2024)\citenamefont {Morvan},
  \citenamefont {Villalonga}, \citenamefont {Mi}, \citenamefont {Mandr{\`a}},
  \citenamefont {Bengtsson}, \citenamefont {Klimov}, \citenamefont {Chen},
  \citenamefont {Hong}, \citenamefont {Erickson}, \citenamefont {Drozdov} \emph
  {et~al.}}]{morvan2024phase}%
  \BibitemOpen
  \bibfield  {author} {\bibinfo {author} {\bibfnamefont {A.}~\bibnamefont
  {Morvan}}, \bibinfo {author} {\bibfnamefont {B.}~\bibnamefont {Villalonga}},
  \bibinfo {author} {\bibfnamefont {X.}~\bibnamefont {Mi}}, \bibinfo {author}
  {\bibfnamefont {S.}~\bibnamefont {Mandr{\`a}}}, \bibinfo {author}
  {\bibfnamefont {A.}~\bibnamefont {Bengtsson}}, \bibinfo {author}
  {\bibfnamefont {P.}~\bibnamefont {Klimov}}, \bibinfo {author} {\bibfnamefont
  {Z.}~\bibnamefont {Chen}}, \bibinfo {author} {\bibfnamefont {S.}~\bibnamefont
  {Hong}}, \bibinfo {author} {\bibfnamefont {C.}~\bibnamefont {Erickson}},
  \bibinfo {author} {\bibfnamefont {I.}~\bibnamefont {Drozdov}}, \emph
  {et~al.},\ }\bibfield  {title} {\bibinfo {title} {Phase transitions in random
  circuit sampling},\ }\href@noop {} {\bibfield  {journal} {\bibinfo  {journal}
  {Nature}\ }\textbf {\bibinfo {volume} {634}},\ \bibinfo {pages} {328}
  (\bibinfo {year} {2024})}\BibitemShut {NoStop}%
\bibitem [{\citenamefont {Gao}\ \emph {et~al.}(2025)\citenamefont {Gao},
  \citenamefont {Fan}, \citenamefont {Zha}, \citenamefont {Bei}, \citenamefont
  {Cai}, \citenamefont {Cai}, \citenamefont {Cao}, \citenamefont {Chen},
  \citenamefont {Chen}, \citenamefont {Chen}, \citenamefont {Chen},
  \citenamefont {Chen}, \citenamefont {Chen}, \citenamefont {Chen},
  \citenamefont {Chen}, \citenamefont {Chu}, \citenamefont {Deng},
  \citenamefont {Deng}, \citenamefont {Ding}, \citenamefont {Ding},
  \citenamefont {Ding}, \citenamefont {Dong} \emph
  {et~al.}}]{gaoEstablishingNewBenchmark2025}%
  \BibitemOpen
  \bibfield  {author} {\bibinfo {author} {\bibfnamefont {D.}~\bibnamefont
  {Gao}}, \bibinfo {author} {\bibfnamefont {D.}~\bibnamefont {Fan}}, \bibinfo
  {author} {\bibfnamefont {C.}~\bibnamefont {Zha}}, \bibinfo {author}
  {\bibfnamefont {J.}~\bibnamefont {Bei}}, \bibinfo {author} {\bibfnamefont
  {G.}~\bibnamefont {Cai}}, \bibinfo {author} {\bibfnamefont {J.}~\bibnamefont
  {Cai}}, \bibinfo {author} {\bibfnamefont {S.}~\bibnamefont {Cao}}, \bibinfo
  {author} {\bibfnamefont {F.}~\bibnamefont {Chen}}, \bibinfo {author}
  {\bibfnamefont {J.}~\bibnamefont {Chen}}, \bibinfo {author} {\bibfnamefont
  {K.}~\bibnamefont {Chen}}, \bibinfo {author} {\bibfnamefont {X.}~\bibnamefont
  {Chen}}, \bibinfo {author} {\bibfnamefont {X.}~\bibnamefont {Chen}}, \bibinfo
  {author} {\bibfnamefont {Z.}~\bibnamefont {Chen}}, \bibinfo {author}
  {\bibfnamefont {Z.}~\bibnamefont {Chen}}, \bibinfo {author} {\bibfnamefont
  {Z.}~\bibnamefont {Chen}}, \bibinfo {author} {\bibfnamefont {W.}~\bibnamefont
  {Chu}}, \bibinfo {author} {\bibfnamefont {H.}~\bibnamefont {Deng}}, \bibinfo
  {author} {\bibfnamefont {Z.}~\bibnamefont {Deng}}, \bibinfo {author}
  {\bibfnamefont {P.}~\bibnamefont {Ding}}, \bibinfo {author} {\bibfnamefont
  {X.}~\bibnamefont {Ding}}, \bibinfo {author} {\bibfnamefont {Z.}~\bibnamefont
  {Ding}}, \bibinfo {author} {\bibfnamefont {S.}~\bibnamefont {Dong}}, \emph
  {et~al.},\ }\bibfield  {title} {\bibinfo {title} {Establishing a {{New
  Benchmark}} in {{Quantum Computational Advantage}} with 105-qubit
  {{Zuchongzhi}} 3.0 {{Processor}}},\ }\href
  {https://doi.org/10.1103/PhysRevLett.134.090601} {\bibfield  {journal}
  {\bibinfo  {journal} {Physical Review Letters}\ }\textbf {\bibinfo {volume}
  {134}},\ \bibinfo {pages} {090601} (\bibinfo {year} {2025})}\BibitemShut
  {NoStop}%
\bibitem [{\citenamefont {Abanin}\ \emph {et~al.}(2025)\citenamefont {Abanin},
  \citenamefont {Acharya}, \citenamefont {Aghababaie-Beni}, \citenamefont
  {Aigeldinger}, \citenamefont {Ajoy}, \citenamefont {Alcaraz}, \citenamefont
  {Aleiner}, \citenamefont {Andersen}, \citenamefont {Ansmann}, \citenamefont
  {Arute}, \citenamefont {Arya}, \citenamefont {Asfaw}, \citenamefont
  {Astrakhantsev}, \citenamefont {Atalaya} \emph
  {et~al.}}]{abanin2025constructiveinterferenceedgequantum}%
  \BibitemOpen
  \bibfield  {author} {\bibinfo {author} {\bibfnamefont {D.~A.}\ \bibnamefont
  {Abanin}}, \bibinfo {author} {\bibfnamefont {R.}~\bibnamefont {Acharya}},
  \bibinfo {author} {\bibfnamefont {L.}~\bibnamefont {Aghababaie-Beni}},
  \bibinfo {author} {\bibfnamefont {G.}~\bibnamefont {Aigeldinger}}, \bibinfo
  {author} {\bibfnamefont {A.}~\bibnamefont {Ajoy}}, \bibinfo {author}
  {\bibfnamefont {R.}~\bibnamefont {Alcaraz}}, \bibinfo {author} {\bibfnamefont
  {I.}~\bibnamefont {Aleiner}}, \bibinfo {author} {\bibfnamefont {T.~I.}\
  \bibnamefont {Andersen}}, \bibinfo {author} {\bibfnamefont {M.}~\bibnamefont
  {Ansmann}}, \bibinfo {author} {\bibfnamefont {F.}~\bibnamefont {Arute}},
  \bibinfo {author} {\bibfnamefont {K.}~\bibnamefont {Arya}}, \bibinfo {author}
  {\bibfnamefont {A.}~\bibnamefont {Asfaw}}, \bibinfo {author} {\bibfnamefont
  {N.}~\bibnamefont {Astrakhantsev}}, \bibinfo {author} {\bibfnamefont
  {J.}~\bibnamefont {Atalaya}}, \emph {et~al.},\ }\href
  {https://arxiv.org/abs/2506.10191} {\bibinfo {title} {Constructive
  interference at the edge of quantum ergodic dynamics}} (\bibinfo {year}
  {2025}),\ \Eprint {https://arxiv.org/abs/2506.10191} {arXiv:2506.10191
  [quant-ph]} \BibitemShut {NoStop}%
\bibitem [{\citenamefont {Aaronson}\ and\ \citenamefont
  {Arkhipov}(2013)}]{aaronson2010computationalcomplexitylinearoptics}%
  \BibitemOpen
  \bibfield  {author} {\bibinfo {author} {\bibfnamefont {S.}~\bibnamefont
  {Aaronson}}\ and\ \bibinfo {author} {\bibfnamefont {A.}~\bibnamefont
  {Arkhipov}},\ }\bibfield  {title} {\bibinfo {title} {The {{Computational
  Complexity}} of {{Linear Optics}}},\ }\href
  {https://doi.org/10.4086/toc.2013.v009a004} {\bibfield  {journal} {\bibinfo
  {journal} {Theory of Computing}\ }\textbf {\bibinfo {volume} {9}},\ \bibinfo
  {pages} {143} (\bibinfo {year} {2013})}\BibitemShut {NoStop}%
\bibitem [{\citenamefont {Bouland}\ \emph {et~al.}(2019)\citenamefont
  {Bouland}, \citenamefont {Fefferman}, \citenamefont {Nirkhe},\ and\
  \citenamefont {Vazirani}}]{bouland2019complexity}%
  \BibitemOpen
  \bibfield  {author} {\bibinfo {author} {\bibfnamefont {A.}~\bibnamefont
  {Bouland}}, \bibinfo {author} {\bibfnamefont {B.}~\bibnamefont {Fefferman}},
  \bibinfo {author} {\bibfnamefont {C.}~\bibnamefont {Nirkhe}},\ and\ \bibinfo
  {author} {\bibfnamefont {U.}~\bibnamefont {Vazirani}},\ }\bibfield  {title}
  {\bibinfo {title} {On the complexity and verification of quantum random
  circuit sampling},\ }\href
  {https://www.nature.com/articles/s41567-018-0318-2} {\bibfield  {journal}
  {\bibinfo  {journal} {Nature Physics}\ }\textbf {\bibinfo {volume} {15}},\
  \bibinfo {pages} {159} (\bibinfo {year} {2019})}\BibitemShut {NoStop}%
\bibitem [{\citenamefont {Pan}\ and\ \citenamefont
  {Zhang}(2022)}]{panSimulationQuantumCircuits2022}%
  \BibitemOpen
  \bibfield  {author} {\bibinfo {author} {\bibfnamefont {F.}~\bibnamefont
  {Pan}}\ and\ \bibinfo {author} {\bibfnamefont {P.}~\bibnamefont {Zhang}},\
  }\bibfield  {title} {\bibinfo {title} {Simulation of {{Quantum Circuits
  Using}} the {{Big-Batch Tensor Network Method}}},\ }\href
  {https://doi.org/10.1103/PhysRevLett.128.030501} {\bibfield  {journal}
  {\bibinfo  {journal} {Physical Review Letters}\ }\textbf {\bibinfo {volume}
  {128}},\ \bibinfo {pages} {030501} (\bibinfo {year} {2022})}\BibitemShut
  {NoStop}%
\bibitem [{\citenamefont {Pan}\ \emph {et~al.}(2022)\citenamefont {Pan},
  \citenamefont {Chen},\ and\ \citenamefont
  {Zhang}}]{panSolvingSamplingProblem2022a}%
  \BibitemOpen
  \bibfield  {author} {\bibinfo {author} {\bibfnamefont {F.}~\bibnamefont
  {Pan}}, \bibinfo {author} {\bibfnamefont {K.}~\bibnamefont {Chen}},\ and\
  \bibinfo {author} {\bibfnamefont {P.}~\bibnamefont {Zhang}},\ }\bibfield
  {title} {\bibinfo {title} {Solving the {{Sampling Problem}} of the {{Sycamore
  Quantum Circuits}}},\ }\href {https://doi.org/10.1103/PhysRevLett.129.090502}
  {\bibfield  {journal} {\bibinfo  {journal} {Physical Review Letters}\
  }\textbf {\bibinfo {volume} {129}},\ \bibinfo {pages} {090502} (\bibinfo
  {year} {2022})}\BibitemShut {NoStop}%
\bibitem [{\citenamefont {Fu}\ \emph {et~al.}(2024)\citenamefont {Fu},
  \citenamefont {Su}, \citenamefont {Zhong}, \citenamefont {Zhao},
  \citenamefont {Zhang}, \citenamefont {Pan}, \citenamefont {Zhang},
  \citenamefont {Zhao}, \citenamefont {Chen}, \citenamefont {Lu}, \citenamefont
  {Pan}, \citenamefont {Pei}, \citenamefont {Zhang},\ and\ \citenamefont
  {Ouyang}}]{fuSurpassingSycamoreAchieving2024}%
  \BibitemOpen
  \bibfield  {author} {\bibinfo {author} {\bibfnamefont {R.}~\bibnamefont
  {Fu}}, \bibinfo {author} {\bibfnamefont {Z.}~\bibnamefont {Su}}, \bibinfo
  {author} {\bibfnamefont {H.-S.}\ \bibnamefont {Zhong}}, \bibinfo {author}
  {\bibfnamefont {X.}~\bibnamefont {Zhao}}, \bibinfo {author} {\bibfnamefont
  {J.}~\bibnamefont {Zhang}}, \bibinfo {author} {\bibfnamefont
  {F.}~\bibnamefont {Pan}}, \bibinfo {author} {\bibfnamefont {P.}~\bibnamefont
  {Zhang}}, \bibinfo {author} {\bibfnamefont {X.}~\bibnamefont {Zhao}},
  \bibinfo {author} {\bibfnamefont {M.-C.}\ \bibnamefont {Chen}}, \bibinfo
  {author} {\bibfnamefont {C.-Y.}\ \bibnamefont {Lu}}, \bibinfo {author}
  {\bibfnamefont {J.-W.}\ \bibnamefont {Pan}}, \bibinfo {author} {\bibfnamefont
  {Z.}~\bibnamefont {Pei}}, \bibinfo {author} {\bibfnamefont {X.}~\bibnamefont
  {Zhang}},\ and\ \bibinfo {author} {\bibfnamefont {W.}~\bibnamefont
  {Ouyang}},\ }\bibfield  {title} {\bibinfo {title} {Surpassing sycamore:
  Achieving energetic superiority through system-level circuit simulation},\
  }in\ \href {https://doi.org/10.1109/SC41406.2024.00085} {\emph {\bibinfo
  {booktitle} {Proceedings of the International Conference for High Performance
  Computing, Networking, Storage, and Analysis}}},\ \bibinfo {series and
  number} {SC '24}\ (\bibinfo  {publisher} {IEEE Press},\ \bibinfo {year}
  {2024})\BibitemShut {NoStop}%
\bibitem [{\citenamefont {Pan}\ \emph {et~al.}(2024)\citenamefont {Pan},
  \citenamefont {Gu}, \citenamefont {Kuang}, \citenamefont {Liu},\ and\
  \citenamefont {Zhang}}]{panEfficientQuantumCircuit2024}%
  \BibitemOpen
  \bibfield  {author} {\bibinfo {author} {\bibfnamefont {F.}~\bibnamefont
  {Pan}}, \bibinfo {author} {\bibfnamefont {H.}~\bibnamefont {Gu}}, \bibinfo
  {author} {\bibfnamefont {L.}~\bibnamefont {Kuang}}, \bibinfo {author}
  {\bibfnamefont {B.}~\bibnamefont {Liu}},\ and\ \bibinfo {author}
  {\bibfnamefont {P.}~\bibnamefont {Zhang}},\ }\bibfield  {title} {\bibinfo
  {title} {Efficient {{Quantum Circuit Simulation}} by {{Tensor Network
  Methods}} on {{Modern GPUs}}},\ }\href {https://doi.org/10.1145/3696465}
  {\bibfield  {journal} {\bibinfo  {journal} {ACM Transactions on Quantum
  Computing}\ }\textbf {\bibinfo {volume} {5}},\ \bibinfo {pages} {26:1}
  (\bibinfo {year} {2024})}\BibitemShut {NoStop}%
\bibitem [{\citenamefont {Zhao}\ \emph {et~al.}(2025)\citenamefont {Zhao},
  \citenamefont {Zhong}, \citenamefont {Pan}, \citenamefont {Chen},
  \citenamefont {Fu}, \citenamefont {Su}, \citenamefont {Xie}, \citenamefont
  {Zhao}, \citenamefont {Zhang}, \citenamefont {Ouyang}, \citenamefont {Lu},
  \citenamefont {Pan},\ and\ \citenamefont
  {Chen}}]{zhaoLeapfroggingSycamoreHarnessing2025}%
  \BibitemOpen
  \bibfield  {author} {\bibinfo {author} {\bibfnamefont {X.-H.}\ \bibnamefont
  {Zhao}}, \bibinfo {author} {\bibfnamefont {H.-S.}\ \bibnamefont {Zhong}},
  \bibinfo {author} {\bibfnamefont {F.}~\bibnamefont {Pan}}, \bibinfo {author}
  {\bibfnamefont {Z.-H.}\ \bibnamefont {Chen}}, \bibinfo {author}
  {\bibfnamefont {R.}~\bibnamefont {Fu}}, \bibinfo {author} {\bibfnamefont
  {Z.}~\bibnamefont {Su}}, \bibinfo {author} {\bibfnamefont {X.}~\bibnamefont
  {Xie}}, \bibinfo {author} {\bibfnamefont {C.}~\bibnamefont {Zhao}}, \bibinfo
  {author} {\bibfnamefont {P.}~\bibnamefont {Zhang}}, \bibinfo {author}
  {\bibfnamefont {W.}~\bibnamefont {Ouyang}}, \bibinfo {author} {\bibfnamefont
  {C.-Y.}\ \bibnamefont {Lu}}, \bibinfo {author} {\bibfnamefont {J.-W.}\
  \bibnamefont {Pan}},\ and\ \bibinfo {author} {\bibfnamefont {M.-C.}\
  \bibnamefont {Chen}},\ }\bibfield  {title} {\bibinfo {title} {Leapfrogging
  {{Sycamore}}: Harnessing 1432 {{GPUs}} for 7× faster quantum random circuit
  sampling},\ }\href {https://doi.org/10.1093/nsr/nwae317} {\bibfield
  {journal} {\bibinfo  {journal} {National Science Review}\ }\textbf {\bibinfo
  {volume} {12}},\ \bibinfo {pages} {nwae317} (\bibinfo {year}
  {2025})}\BibitemShut {NoStop}%
\bibitem [{\citenamefont {Qi}\ \emph {et~al.}(2020)\citenamefont {Qi},
  \citenamefont {Brod}, \citenamefont {Quesada},\ and\ \citenamefont
  {Garc\'{\i}a-Patr\'on}}]{PhysRevLett.124.100502}%
  \BibitemOpen
  \bibfield  {author} {\bibinfo {author} {\bibfnamefont {H.}~\bibnamefont
  {Qi}}, \bibinfo {author} {\bibfnamefont {D.~J.}\ \bibnamefont {Brod}},
  \bibinfo {author} {\bibfnamefont {N.}~\bibnamefont {Quesada}},\ and\ \bibinfo
  {author} {\bibfnamefont {R.}~\bibnamefont {Garc\'{\i}a-Patr\'on}},\
  }\bibfield  {title} {\bibinfo {title} {Regimes of classical simulability for
  noisy gaussian boson sampling},\ }\href
  {https://doi.org/10.1103/PhysRevLett.124.100502} {\bibfield  {journal}
  {\bibinfo  {journal} {Physical Review Letters}\ }\textbf {\bibinfo {volume}
  {124}},\ \bibinfo {pages} {100502} (\bibinfo {year} {2020})}\BibitemShut
  {NoStop}%
\bibitem [{\citenamefont {Villalonga}\ \emph {et~al.}(2021)\citenamefont
  {Villalonga}, \citenamefont {Niu}, \citenamefont {Li}, \citenamefont {Neven},
  \citenamefont {Platt}, \citenamefont {Smelyanskiy},\ and\ \citenamefont
  {Boixo}}]{villalonga2021efficient}%
  \BibitemOpen
  \bibfield  {author} {\bibinfo {author} {\bibfnamefont {B.}~\bibnamefont
  {Villalonga}}, \bibinfo {author} {\bibfnamefont {M.~Y.}\ \bibnamefont {Niu}},
  \bibinfo {author} {\bibfnamefont {L.}~\bibnamefont {Li}}, \bibinfo {author}
  {\bibfnamefont {H.}~\bibnamefont {Neven}}, \bibinfo {author} {\bibfnamefont
  {J.~C.}\ \bibnamefont {Platt}}, \bibinfo {author} {\bibfnamefont {V.~N.}\
  \bibnamefont {Smelyanskiy}},\ and\ \bibinfo {author} {\bibfnamefont
  {S.}~\bibnamefont {Boixo}},\ }\bibfield  {title} {\bibinfo {title} {Efficient
  approximation of experimental gaussian boson sampling},\ }\href
  {https://arxiv.org/abs/2109.11525} {\bibfield  {journal} {\bibinfo  {journal}
  {arXiv preprint arXiv:2109.11525}\ } (\bibinfo {year} {2021})}\BibitemShut
  {NoStop}%
\bibitem [{\citenamefont {Quesada}\ \emph {et~al.}(2022)\citenamefont
  {Quesada}, \citenamefont {Chadwick}, \citenamefont {Bell}, \citenamefont
  {Arrazola}, \citenamefont {Vincent}, \citenamefont {Qi},\ and\ \citenamefont
  {Garc{\'\i}a-Patr{\'o}n}}]{quesada2022quadratic}%
  \BibitemOpen
  \bibfield  {author} {\bibinfo {author} {\bibfnamefont {N.}~\bibnamefont
  {Quesada}}, \bibinfo {author} {\bibfnamefont {R.~S.}\ \bibnamefont
  {Chadwick}}, \bibinfo {author} {\bibfnamefont {B.~A.}\ \bibnamefont {Bell}},
  \bibinfo {author} {\bibfnamefont {J.~M.}\ \bibnamefont {Arrazola}}, \bibinfo
  {author} {\bibfnamefont {T.}~\bibnamefont {Vincent}}, \bibinfo {author}
  {\bibfnamefont {H.}~\bibnamefont {Qi}},\ and\ \bibinfo {author}
  {\bibfnamefont {R.}~\bibnamefont {Garc{\'\i}a-Patr{\'o}n}},\ }\bibfield
  {title} {\bibinfo {title} {Quadratic speed-up for simulating gaussian boson
  sampling},\ }\href {https://doi.org/10.1103/PRXQuantum.3.010306} {\bibfield
  {journal} {\bibinfo  {journal} {PRX Quantum}\ }\textbf {\bibinfo {volume}
  {3}},\ \bibinfo {pages} {010306} (\bibinfo {year} {2022})}\BibitemShut
  {NoStop}%
\bibitem [{\citenamefont {Bulmer}\ \emph {et~al.}(2022)\citenamefont {Bulmer},
  \citenamefont {Bell}, \citenamefont {Chadwick}, \citenamefont {Jones},
  \citenamefont {Moise}, \citenamefont {Rigazzi}, \citenamefont {Thorbecke},
  \citenamefont {Haus}, \citenamefont {Van~Vaerenbergh}, \citenamefont {Patel}
  \emph {et~al.}}]{bulmer2022boundary}%
  \BibitemOpen
  \bibfield  {author} {\bibinfo {author} {\bibfnamefont {J.~F.}\ \bibnamefont
  {Bulmer}}, \bibinfo {author} {\bibfnamefont {B.~A.}\ \bibnamefont {Bell}},
  \bibinfo {author} {\bibfnamefont {R.~S.}\ \bibnamefont {Chadwick}}, \bibinfo
  {author} {\bibfnamefont {A.~E.}\ \bibnamefont {Jones}}, \bibinfo {author}
  {\bibfnamefont {D.}~\bibnamefont {Moise}}, \bibinfo {author} {\bibfnamefont
  {A.}~\bibnamefont {Rigazzi}}, \bibinfo {author} {\bibfnamefont
  {J.}~\bibnamefont {Thorbecke}}, \bibinfo {author} {\bibfnamefont {U.-U.}\
  \bibnamefont {Haus}}, \bibinfo {author} {\bibfnamefont {T.}~\bibnamefont
  {Van~Vaerenbergh}}, \bibinfo {author} {\bibfnamefont {R.~B.}\ \bibnamefont
  {Patel}}, \emph {et~al.},\ }\bibfield  {title} {\bibinfo {title} {The
  boundary for quantum advantage in gaussian boson sampling},\ }\href
  {https://www.science.org/doi/10.1126/sciadv.abl9236} {\bibfield  {journal}
  {\bibinfo  {journal} {Science advances}\ }\textbf {\bibinfo {volume} {8}},\
  \bibinfo {pages} {eabl9236} (\bibinfo {year} {2022})}\BibitemShut {NoStop}%
\bibitem [{\citenamefont {Oh}\ \emph {et~al.}(2022)\citenamefont {Oh},
  \citenamefont {Lim}, \citenamefont {Fefferman},\ and\ \citenamefont
  {Jiang}}]{oh2022classical}%
  \BibitemOpen
  \bibfield  {author} {\bibinfo {author} {\bibfnamefont {C.}~\bibnamefont
  {Oh}}, \bibinfo {author} {\bibfnamefont {Y.}~\bibnamefont {Lim}}, \bibinfo
  {author} {\bibfnamefont {B.}~\bibnamefont {Fefferman}},\ and\ \bibinfo
  {author} {\bibfnamefont {L.}~\bibnamefont {Jiang}},\ }\bibfield  {title}
  {\bibinfo {title} {Classical simulation of boson sampling based on graph
  structure},\ }\href {https://doi.org/10.1103/PhysRevLett.128.190501}
  {\bibfield  {journal} {\bibinfo  {journal} {Physical Review Letters}\
  }\textbf {\bibinfo {volume} {128}},\ \bibinfo {pages} {190501} (\bibinfo
  {year} {2022})}\BibitemShut {NoStop}%
\bibitem [{\citenamefont {Oh}\ \emph {et~al.}(2023)\citenamefont {Oh},
  \citenamefont {Jiang},\ and\ \citenamefont {Fefferman}}]{oh2023spoofing}%
  \BibitemOpen
  \bibfield  {author} {\bibinfo {author} {\bibfnamefont {C.}~\bibnamefont
  {Oh}}, \bibinfo {author} {\bibfnamefont {L.}~\bibnamefont {Jiang}},\ and\
  \bibinfo {author} {\bibfnamefont {B.}~\bibnamefont {Fefferman}},\ }\bibfield
  {title} {\bibinfo {title} {Spoofing cross-entropy measure in boson
  sampling},\ }\href {https://doi.org/10.1103/PhysRevLett.131.010401}
  {\bibfield  {journal} {\bibinfo  {journal} {Physical Review Letters}\
  }\textbf {\bibinfo {volume} {131}},\ \bibinfo {pages} {010401} (\bibinfo
  {year} {2023})}\BibitemShut {NoStop}%
\bibitem [{\citenamefont {Oh}\ \emph {et~al.}(2024)\citenamefont {Oh},
  \citenamefont {Liu}, \citenamefont {Alexeev}, \citenamefont {Fefferman},\
  and\ \citenamefont {Jiang}}]{oh2024classical}%
  \BibitemOpen
  \bibfield  {author} {\bibinfo {author} {\bibfnamefont {C.}~\bibnamefont
  {Oh}}, \bibinfo {author} {\bibfnamefont {M.}~\bibnamefont {Liu}}, \bibinfo
  {author} {\bibfnamefont {Y.}~\bibnamefont {Alexeev}}, \bibinfo {author}
  {\bibfnamefont {B.}~\bibnamefont {Fefferman}},\ and\ \bibinfo {author}
  {\bibfnamefont {L.}~\bibnamefont {Jiang}},\ }\bibfield  {title} {\bibinfo
  {title} {Classical algorithm for simulating experimental gaussian boson
  sampling},\ }\href {https://www.nature.com/articles/s41567-024-02535-8}
  {\bibfield  {journal} {\bibinfo  {journal} {Nature Physics}\ }\textbf
  {\bibinfo {volume} {20}},\ \bibinfo {pages} {1461} (\bibinfo {year}
  {2024})}\BibitemShut {NoStop}%
\bibitem [{\citenamefont {Hamilton}\ \emph {et~al.}(2017)\citenamefont
  {Hamilton}, \citenamefont {Kruse}, \citenamefont {Sansoni}, \citenamefont
  {Barkhofen}, \citenamefont {Silberhorn},\ and\ \citenamefont
  {Jex}}]{hamilton2017gaussian}%
  \BibitemOpen
  \bibfield  {author} {\bibinfo {author} {\bibfnamefont {C.~S.}\ \bibnamefont
  {Hamilton}}, \bibinfo {author} {\bibfnamefont {R.}~\bibnamefont {Kruse}},
  \bibinfo {author} {\bibfnamefont {L.}~\bibnamefont {Sansoni}}, \bibinfo
  {author} {\bibfnamefont {S.}~\bibnamefont {Barkhofen}}, \bibinfo {author}
  {\bibfnamefont {C.}~\bibnamefont {Silberhorn}},\ and\ \bibinfo {author}
  {\bibfnamefont {I.}~\bibnamefont {Jex}},\ }\bibfield  {title} {\bibinfo
  {title} {Gaussian boson sampling},\ }\href
  {https://doi.org/10.1103/PhysRevLett.119.170501} {\bibfield  {journal}
  {\bibinfo  {journal} {Physical review letters}\ }\textbf {\bibinfo {volume}
  {119}},\ \bibinfo {pages} {170501} (\bibinfo {year} {2017})}\BibitemShut
  {NoStop}%
\bibitem [{\citenamefont {Quesada}\ \emph {et~al.}(2018)\citenamefont
  {Quesada}, \citenamefont {Arrazola},\ and\ \citenamefont
  {Killoran}}]{quesada2018gaussian}%
  \BibitemOpen
  \bibfield  {author} {\bibinfo {author} {\bibfnamefont {N.}~\bibnamefont
  {Quesada}}, \bibinfo {author} {\bibfnamefont {J.~M.}\ \bibnamefont
  {Arrazola}},\ and\ \bibinfo {author} {\bibfnamefont {N.}~\bibnamefont
  {Killoran}},\ }\bibfield  {title} {\bibinfo {title} {Gaussian boson sampling
  using threshold detectors},\ }\href
  {https://doi.org/10.1103/PhysRevA.98.062322} {\bibfield  {journal} {\bibinfo
  {journal} {Physical Review A}\ }\textbf {\bibinfo {volume} {98}},\ \bibinfo
  {pages} {062322} (\bibinfo {year} {2018})}\BibitemShut {NoStop}%
\bibitem [{\citenamefont {Arrazola}\ \emph {et~al.}(2018)\citenamefont
  {Arrazola}, \citenamefont {Bromley},\ and\ \citenamefont
  {Rebentrost}}]{arrazolaQuantumApproximateOptimization2018}%
  \BibitemOpen
  \bibfield  {author} {\bibinfo {author} {\bibfnamefont {J.~M.}\ \bibnamefont
  {Arrazola}}, \bibinfo {author} {\bibfnamefont {T.~R.}\ \bibnamefont
  {Bromley}},\ and\ \bibinfo {author} {\bibfnamefont {P.}~\bibnamefont
  {Rebentrost}},\ }\bibfield  {title} {\bibinfo {title} {Quantum approximate
  optimization with {{Gaussian}} boson sampling},\ }\href
  {https://doi.org/10.1103/PhysRevA.98.012322} {\bibfield  {journal} {\bibinfo
  {journal} {Physical Review A}\ }\textbf {\bibinfo {volume} {98}},\ \bibinfo
  {pages} {012322} (\bibinfo {year} {2018})}\BibitemShut {NoStop}%
\bibitem [{\citenamefont {Arrazola}\ and\ \citenamefont
  {Bromley}(2018)}]{arrazolaUsingGaussianBoson2018}%
  \BibitemOpen
  \bibfield  {author} {\bibinfo {author} {\bibfnamefont {J.~M.}\ \bibnamefont
  {Arrazola}}\ and\ \bibinfo {author} {\bibfnamefont {T.~R.}\ \bibnamefont
  {Bromley}},\ }\bibfield  {title} {\bibinfo {title} {Using {{Gaussian Boson
  Sampling}} to {{Find Dense Subgraphs}}},\ }\href
  {https://doi.org/10.1103/PhysRevLett.121.030503} {\bibfield  {journal}
  {\bibinfo  {journal} {Physical Review Letters}\ }\textbf {\bibinfo {volume}
  {121}},\ \bibinfo {pages} {030503} (\bibinfo {year} {2018})}\BibitemShut
  {NoStop}%
\bibitem [{\citenamefont {Banchi}\ \emph {et~al.}(2020)\citenamefont {Banchi},
  \citenamefont {Fingerhuth}, \citenamefont {Babej}, \citenamefont {Ing},\ and\
  \citenamefont {Arrazola}}]{banchiMolecularDockingGaussian}%
  \BibitemOpen
  \bibfield  {author} {\bibinfo {author} {\bibfnamefont {L.}~\bibnamefont
  {Banchi}}, \bibinfo {author} {\bibfnamefont {M.}~\bibnamefont {Fingerhuth}},
  \bibinfo {author} {\bibfnamefont {T.}~\bibnamefont {Babej}}, \bibinfo
  {author} {\bibfnamefont {C.}~\bibnamefont {Ing}},\ and\ \bibinfo {author}
  {\bibfnamefont {J.~M.}\ \bibnamefont {Arrazola}},\ }\bibfield  {title}
  {\bibinfo {title} {Molecular docking with {{Gaussian Boson Sampling}}},\
  }\href {https://doi.org/10.1126/sciadv.aax1950} {\bibfield  {journal}
  {\bibinfo  {journal} {Science Advances}\ }\textbf {\bibinfo {volume} {6}},\
  \bibinfo {pages} {eaax1950} (\bibinfo {year} {2020})}\BibitemShut {NoStop}%
\bibitem [{\citenamefont {Huh}\ \emph {et~al.}(2015)\citenamefont {Huh},
  \citenamefont {Guerreschi}, \citenamefont {Peropadre}, \citenamefont
  {McClean},\ and\ \citenamefont
  {{Aspuru-Guzik}}}]{huhBosonSamplingMolecular2015}%
  \BibitemOpen
  \bibfield  {author} {\bibinfo {author} {\bibfnamefont {J.}~\bibnamefont
  {Huh}}, \bibinfo {author} {\bibfnamefont {G.~G.}\ \bibnamefont {Guerreschi}},
  \bibinfo {author} {\bibfnamefont {B.}~\bibnamefont {Peropadre}}, \bibinfo
  {author} {\bibfnamefont {J.~R.}\ \bibnamefont {McClean}},\ and\ \bibinfo
  {author} {\bibfnamefont {A.}~\bibnamefont {{Aspuru-Guzik}}},\ }\bibfield
  {title} {\bibinfo {title} {Boson sampling for molecular vibronic spectra},\
  }\href {https://doi.org/10.1038/nphoton.2015.153} {\bibfield  {journal}
  {\bibinfo  {journal} {Nature Photonics}\ }\textbf {\bibinfo {volume} {9}},\
  \bibinfo {pages} {615} (\bibinfo {year} {2015})}\BibitemShut {NoStop}%
\bibitem [{\citenamefont {Jahangiri}\ \emph {et~al.}(2020)\citenamefont
  {Jahangiri}, \citenamefont {Miguel~Arrazola}, \citenamefont {Quesada},\ and\
  \citenamefont {Delgado}}]{jahangiriQuantumAlgorithmSimulating2020}%
  \BibitemOpen
  \bibfield  {author} {\bibinfo {author} {\bibfnamefont {S.}~\bibnamefont
  {Jahangiri}}, \bibinfo {author} {\bibfnamefont {J.}~\bibnamefont
  {Miguel~Arrazola}}, \bibinfo {author} {\bibfnamefont {N.}~\bibnamefont
  {Quesada}},\ and\ \bibinfo {author} {\bibfnamefont {A.}~\bibnamefont
  {Delgado}},\ }\bibfield  {title} {\bibinfo {title} {Quantum algorithm for
  simulating molecular vibrational excitations},\ }\href
  {https://doi.org/10.1039/D0CP03593A} {\bibfield  {journal} {\bibinfo
  {journal} {Physical Chemistry Chemical Physics}\ }\textbf {\bibinfo {volume}
  {22}},\ \bibinfo {pages} {25528} (\bibinfo {year} {2020})}\BibitemShut
  {NoStop}%
\bibitem [{\citenamefont {Jahangiri}\ \emph {et~al.}(2021)\citenamefont
  {Jahangiri}, \citenamefont {Arrazola},\ and\ \citenamefont
  {Delgado}}]{jahangiriQuantumAlgorithmSimulating2021}%
  \BibitemOpen
  \bibfield  {author} {\bibinfo {author} {\bibfnamefont {S.}~\bibnamefont
  {Jahangiri}}, \bibinfo {author} {\bibfnamefont {J.~M.}\ \bibnamefont
  {Arrazola}},\ and\ \bibinfo {author} {\bibfnamefont {A.}~\bibnamefont
  {Delgado}},\ }\bibfield  {title} {\bibinfo {title} {Quantum {{Algorithm}} for
  {{Simulating Single-Molecule Electron Transport}}},\ }\href
  {https://doi.org/10.1021/acs.jpclett.0c03724} {\bibfield  {journal} {\bibinfo
   {journal} {The Journal of Physical Chemistry Letters}\ }\textbf {\bibinfo
  {volume} {12}},\ \bibinfo {pages} {1256} (\bibinfo {year}
  {2021})}\BibitemShut {NoStop}%
\bibitem [{\citenamefont {Deng}\ \emph
  {et~al.}(2023{\natexlab{b}})\citenamefont {Deng}, \citenamefont {Gong},
  \citenamefont {Gu}, \citenamefont {Zhang}, \citenamefont {Liu}, \citenamefont
  {Su}, \citenamefont {Tang}, \citenamefont {Xu}, \citenamefont {Jia},
  \citenamefont {Chen}, \citenamefont {Zhong}, \citenamefont {Wang},
  \citenamefont {Yan}, \citenamefont {Hu}, \citenamefont {Huang}, \citenamefont
  {Zhang}, \citenamefont {Li}, \citenamefont {Jiang}, \citenamefont {You},
  \citenamefont {Wang}, \citenamefont {Li}, \citenamefont {Liu}, \citenamefont
  {Lu},\ and\ \citenamefont {Pan}}]{dengSolvingGraphProblems2023a}%
  \BibitemOpen
  \bibfield  {author} {\bibinfo {author} {\bibfnamefont {Y.-H.}\ \bibnamefont
  {Deng}}, \bibinfo {author} {\bibfnamefont {S.-Q.}\ \bibnamefont {Gong}},
  \bibinfo {author} {\bibfnamefont {Y.-C.}\ \bibnamefont {Gu}}, \bibinfo
  {author} {\bibfnamefont {Z.-J.}\ \bibnamefont {Zhang}}, \bibinfo {author}
  {\bibfnamefont {H.-L.}\ \bibnamefont {Liu}}, \bibinfo {author} {\bibfnamefont
  {H.}~\bibnamefont {Su}}, \bibinfo {author} {\bibfnamefont {H.-Y.}\
  \bibnamefont {Tang}}, \bibinfo {author} {\bibfnamefont {J.-M.}\ \bibnamefont
  {Xu}}, \bibinfo {author} {\bibfnamefont {M.-H.}\ \bibnamefont {Jia}},
  \bibinfo {author} {\bibfnamefont {M.-C.}\ \bibnamefont {Chen}}, \bibinfo
  {author} {\bibfnamefont {H.-S.}\ \bibnamefont {Zhong}}, \bibinfo {author}
  {\bibfnamefont {H.}~\bibnamefont {Wang}}, \bibinfo {author} {\bibfnamefont
  {J.}~\bibnamefont {Yan}}, \bibinfo {author} {\bibfnamefont {Y.}~\bibnamefont
  {Hu}}, \bibinfo {author} {\bibfnamefont {J.}~\bibnamefont {Huang}}, \bibinfo
  {author} {\bibfnamefont {W.-J.}\ \bibnamefont {Zhang}}, \bibinfo {author}
  {\bibfnamefont {H.}~\bibnamefont {Li}}, \bibinfo {author} {\bibfnamefont
  {X.}~\bibnamefont {Jiang}}, \bibinfo {author} {\bibfnamefont
  {L.}~\bibnamefont {You}}, \bibinfo {author} {\bibfnamefont {Z.}~\bibnamefont
  {Wang}}, \bibinfo {author} {\bibfnamefont {L.}~\bibnamefont {Li}}, \bibinfo
  {author} {\bibfnamefont {N.-L.}\ \bibnamefont {Liu}}, \bibinfo {author}
  {\bibfnamefont {C.-Y.}\ \bibnamefont {Lu}},\ and\ \bibinfo {author}
  {\bibfnamefont {J.-W.}\ \bibnamefont {Pan}},\ }\bibfield  {title} {\bibinfo
  {title} {Solving {{Graph Problems Using Gaussian Boson Sampling}}},\
  }\bibfield  {journal} {\bibinfo  {journal} {Physical Review Letters}\
  }\textbf {\bibinfo {volume} {130}},\ \href
  {https://doi.org/10.1103/physrevlett.130.190601}
  {10.1103/physrevlett.130.190601} (\bibinfo {year}
  {2023}{\natexlab{b}})\BibitemShut {NoStop}%
\bibitem [{\citenamefont {Shang}\ \emph {et~al.}(2024)\citenamefont {Shang},
  \citenamefont {Zhong}, \citenamefont {Zhang}, \citenamefont {Yu},
  \citenamefont {Yuan}, \citenamefont {Lu}, \citenamefont {Pan},\ and\
  \citenamefont {Chen}}]{shangBosonSamplingEnhanced2024a}%
  \BibitemOpen
  \bibfield  {author} {\bibinfo {author} {\bibfnamefont {Z.-X.}\ \bibnamefont
  {Shang}}, \bibinfo {author} {\bibfnamefont {H.-S.}\ \bibnamefont {Zhong}},
  \bibinfo {author} {\bibfnamefont {Y.-K.}\ \bibnamefont {Zhang}}, \bibinfo
  {author} {\bibfnamefont {C.-C.}\ \bibnamefont {Yu}}, \bibinfo {author}
  {\bibfnamefont {X.}~\bibnamefont {Yuan}}, \bibinfo {author} {\bibfnamefont
  {C.-Y.}\ \bibnamefont {Lu}}, \bibinfo {author} {\bibfnamefont {J.-W.}\
  \bibnamefont {Pan}},\ and\ \bibinfo {author} {\bibfnamefont {M.-C.}\
  \bibnamefont {Chen}},\ }\href {https://doi.org/10.48550/arXiv.2403.16698}
  {\bibinfo {title} {Boson sampling enhanced quantum chemistry}} (\bibinfo
  {year} {2024}),\ \Eprint {https://arxiv.org/abs/2403.16698} {arXiv:2403.16698
  [quant-ph]} \BibitemShut {NoStop}%
\bibitem [{\citenamefont {Cimini}\ \emph {et~al.}(2025)\citenamefont {Cimini},
  \citenamefont {Sohoni}, \citenamefont {Presutti}, \citenamefont {Malia},
  \citenamefont {Ma}, \citenamefont {Yanagimoto}, \citenamefont {Wang},
  \citenamefont {Onodera}, \citenamefont {Wright},\ and\ \citenamefont
  {McMahon}}]{cimini2025largescalequantumreservoircomputing}%
  \BibitemOpen
  \bibfield  {author} {\bibinfo {author} {\bibfnamefont {V.}~\bibnamefont
  {Cimini}}, \bibinfo {author} {\bibfnamefont {M.~M.}\ \bibnamefont {Sohoni}},
  \bibinfo {author} {\bibfnamefont {F.}~\bibnamefont {Presutti}}, \bibinfo
  {author} {\bibfnamefont {B.~K.}\ \bibnamefont {Malia}}, \bibinfo {author}
  {\bibfnamefont {S.-Y.}\ \bibnamefont {Ma}}, \bibinfo {author} {\bibfnamefont
  {R.}~\bibnamefont {Yanagimoto}}, \bibinfo {author} {\bibfnamefont
  {T.}~\bibnamefont {Wang}}, \bibinfo {author} {\bibfnamefont {T.}~\bibnamefont
  {Onodera}}, \bibinfo {author} {\bibfnamefont {L.~G.}\ \bibnamefont
  {Wright}},\ and\ \bibinfo {author} {\bibfnamefont {P.~L.}\ \bibnamefont
  {McMahon}},\ }\href {https://arxiv.org/abs/2505.13695} {\bibinfo {title}
  {Large-scale quantum reservoir computing using a gaussian boson sampler}}
  (\bibinfo {year} {2025}),\ \Eprint {https://arxiv.org/abs/2505.13695}
  {arXiv:2505.13695 [quant-ph]} \BibitemShut {NoStop}%
\bibitem [{\citenamefont {Gong}\ \emph {et~al.}(2025)\citenamefont {Gong},
  \citenamefont {Chen}, \citenamefont {Liu}, \citenamefont {Su}, \citenamefont
  {Gu}, \citenamefont {Tang}, \citenamefont {Jia}, \citenamefont {Deng},
  \citenamefont {Wei}, \citenamefont {Wang}, \citenamefont {Zhong},
  \citenamefont {Jiang}, \citenamefont {Li}, \citenamefont {Liu}, \citenamefont
  {Lu},\ and\ \citenamefont {Pan}}]{gong2025enhancedimagerecognitionusing}%
  \BibitemOpen
  \bibfield  {author} {\bibinfo {author} {\bibfnamefont {S.-Q.}\ \bibnamefont
  {Gong}}, \bibinfo {author} {\bibfnamefont {M.-C.}\ \bibnamefont {Chen}},
  \bibinfo {author} {\bibfnamefont {H.-L.}\ \bibnamefont {Liu}}, \bibinfo
  {author} {\bibfnamefont {H.}~\bibnamefont {Su}}, \bibinfo {author}
  {\bibfnamefont {Y.-C.}\ \bibnamefont {Gu}}, \bibinfo {author} {\bibfnamefont
  {H.-Y.}\ \bibnamefont {Tang}}, \bibinfo {author} {\bibfnamefont {M.-H.}\
  \bibnamefont {Jia}}, \bibinfo {author} {\bibfnamefont {Y.-H.}\ \bibnamefont
  {Deng}}, \bibinfo {author} {\bibfnamefont {Q.}~\bibnamefont {Wei}}, \bibinfo
  {author} {\bibfnamefont {H.}~\bibnamefont {Wang}}, \bibinfo {author}
  {\bibfnamefont {H.-S.}\ \bibnamefont {Zhong}}, \bibinfo {author}
  {\bibfnamefont {X.}~\bibnamefont {Jiang}}, \bibinfo {author} {\bibfnamefont
  {L.}~\bibnamefont {Li}}, \bibinfo {author} {\bibfnamefont {N.-L.}\
  \bibnamefont {Liu}}, \bibinfo {author} {\bibfnamefont {C.-Y.}\ \bibnamefont
  {Lu}},\ and\ \bibinfo {author} {\bibfnamefont {J.-W.}\ \bibnamefont {Pan}},\
  }\href {https://arxiv.org/abs/2506.19707} {\bibinfo {title} {Enhanced image
  recognition using gaussian boson sampling}} (\bibinfo {year} {2025}),\
  \Eprint {https://arxiv.org/abs/2506.19707} {arXiv:2506.19707 [quant-ph]}
  \BibitemShut {NoStop}%
\bibitem [{\citenamefont {Tzitrin}\ \emph {et~al.}(2020)\citenamefont
  {Tzitrin}, \citenamefont {Bourassa}, \citenamefont {Menicucci},\ and\
  \citenamefont {Sabapathy}}]{PhysRevA.101.032315}%
  \BibitemOpen
  \bibfield  {author} {\bibinfo {author} {\bibfnamefont {I.}~\bibnamefont
  {Tzitrin}}, \bibinfo {author} {\bibfnamefont {J.~E.}\ \bibnamefont
  {Bourassa}}, \bibinfo {author} {\bibfnamefont {N.~C.}\ \bibnamefont
  {Menicucci}},\ and\ \bibinfo {author} {\bibfnamefont {K.~K.}\ \bibnamefont
  {Sabapathy}},\ }\bibfield  {title} {\bibinfo {title} {Progress towards
  practical qubit computation using approximate gottesman-kitaev-preskill
  codes},\ }\href {https://doi.org/10.1103/PhysRevA.101.032315} {\bibfield
  {journal} {\bibinfo  {journal} {Phys. Rev. A}\ }\textbf {\bibinfo {volume}
  {101}},\ \bibinfo {pages} {032315} (\bibinfo {year} {2020})}\BibitemShut
  {NoStop}%
\bibitem [{\citenamefont {Larsen}\ \emph {et~al.}(2025)\citenamefont {Larsen},
  \citenamefont {Bourassa}, \citenamefont {Kocsis}, \citenamefont {Tasker},
  \citenamefont {Chadwick}, \citenamefont {Gonz{\'a}lez-Arciniegas},
  \citenamefont {Hastrup}, \citenamefont {Lopetegui-Gonz{\'a}lez},
  \citenamefont {Miatto}, \citenamefont {Motamedi} \emph
  {et~al.}}]{larsen2025integrated}%
  \BibitemOpen
  \bibfield  {author} {\bibinfo {author} {\bibfnamefont {M.}~\bibnamefont
  {Larsen}}, \bibinfo {author} {\bibfnamefont {J.}~\bibnamefont {Bourassa}},
  \bibinfo {author} {\bibfnamefont {S.}~\bibnamefont {Kocsis}}, \bibinfo
  {author} {\bibfnamefont {J.}~\bibnamefont {Tasker}}, \bibinfo {author}
  {\bibfnamefont {R.}~\bibnamefont {Chadwick}}, \bibinfo {author}
  {\bibfnamefont {C.}~\bibnamefont {Gonz{\'a}lez-Arciniegas}}, \bibinfo
  {author} {\bibfnamefont {J.}~\bibnamefont {Hastrup}}, \bibinfo {author}
  {\bibfnamefont {C.}~\bibnamefont {Lopetegui-Gonz{\'a}lez}}, \bibinfo {author}
  {\bibfnamefont {F.}~\bibnamefont {Miatto}}, \bibinfo {author} {\bibfnamefont
  {A.}~\bibnamefont {Motamedi}}, \emph {et~al.},\ }\bibfield  {title} {\bibinfo
  {title} {Integrated photonic source of gottesman--kitaev--preskill qubits},\
  }\href {https://www.nature.com/articles/s41586-025-09044-5} {\bibfield
  {journal} {\bibinfo  {journal} {Nature}\ ,\ \bibinfo {pages} {1}} (\bibinfo
  {year} {2025})}\BibitemShut {NoStop}%
\bibitem [{Chi(2020)}]{ChineseBosonSamplingExperiment2020}%
  \BibitemOpen
  \href {https://scottaaronson.blog/?p=5159} {\bibinfo {title} {Chinese
  {{BosonSampling}} experiment: The gloves are off}},\ \bibinfo {howpublished}
  {https://scottaaronson.blog/?p=5159} (\bibinfo {year} {2020})\BibitemShut
  {NoStop}%
\bibitem [{xai()}]{xai}%
  \BibitemOpen
  \href {https://x.ai/colossus} {\bibinfo {title} {Xai-colossus}},\ \bibinfo
  {howpublished} {https://x.ai/colossus}\BibitemShut {NoStop}%
\bibitem [{Nov()}]{November2024TOP500}%
  \BibitemOpen
  \href {https://www.top500.org/lists/top500/2024/11/} {\bibinfo {title}
  {November 2024 {\textbar} {{TOP500}}}},\ \bibinfo {howpublished}
  {https://www.top500.org/lists/top500/2024/11/}\BibitemShut {NoStop}%
\bibitem [{\citenamefont {Chen}\ \emph {et~al.}()\citenamefont {Chen},
  \citenamefont {Gong}, \citenamefont {Gan}, \citenamefont {Liu}, \citenamefont
  {Yang}, \citenamefont {Lu},\ and\ \citenamefont {Yang}}]{fastMPS}%
  \BibitemOpen
  \bibfield  {author} {\bibinfo {author} {\bibfnamefont {Y.}~\bibnamefont
  {Chen}}, \bibinfo {author} {\bibfnamefont {S.-Q.}\ \bibnamefont {Gong}},
  \bibinfo {author} {\bibfnamefont {L.}~\bibnamefont {Gan}}, \bibinfo {author}
  {\bibfnamefont {Y.}~\bibnamefont {Liu}}, \bibinfo {author} {\bibfnamefont
  {A.}~\bibnamefont {Yang}}, \bibinfo {author} {\bibfnamefont {C.-Y.}\
  \bibnamefont {Lu}},\ and\ \bibinfo {author} {\bibfnamefont {G.}~\bibnamefont
  {Yang}},\ }\bibinfo {title} {{FastMPS: Breaking the Scaling Barrier of
  MPS-based Gaussian Boson Sampling Simulation}}\BibitemShut {NoStop}%
\bibitem [{raw()}]{rawData}%
  \BibitemOpen
\bibfield  {title} {  }\href {https://quantum.ustc.edu.cn/web/node/1227}
  {\bibinfo {title} {Raw data of jiuzhang 4.0}},\ \bibinfo {howpublished}
  {https://quantum.ustc.edu.cn/web/node/1227}\BibitemShut {NoStop}%
\end{thebibliography}%

 \end{document}